\PassOptionsToPackage{table}{xcolor}
\documentclass[onefignum,onetabnum,nohypdvips]{siamart171218}

\usepackage{mathrsfs,amsmath,amssymb,bm,latexsym}
\usepackage{lipsum}
\usepackage{amsfonts}
\usepackage{caption, graphicx, subfigure}
\usepackage{epstopdf}
\usepackage{makecell, rotating, bbding}
\usepackage{multirow}
\usepackage{threeparttable}
\usepackage{algorithmic, algorithm}
\usepackage[draft]{changes}
\ifpdf
  \DeclareGraphicsExtensions{.eps,.pdf,.png,.jpg}
\else
  \DeclareGraphicsExtensions{.eps}
\fi

\usepackage{enumitem}
\setlist[enumerate]{leftmargin=.5in}
\setlist[itemize]{leftmargin=.5in}

\newsiamremark{remark}{Remark}
\newsiamremark{hypothesis}{Hypothesis}
\crefname{hypothesis}{Hypothesis}{Hypotheses}
\newsiamthm{claim}{Claim}
\newsiamremark{defy}{Definition}

\headers{Tilt hexagonal GBs: a spectral viewpoint}{
	K. Jiang, W. Si, and J. Xu}

\title{Tilt grain boundaries of hexagonal structures: a spectral viewpoint
	}
\author{Kai Jiang\thanks{
	School of Mathematics and Computational Science, 
	Hunan Key Laboratory for Computation and Simulation in Science and Engineering,
	Xiangtan University, Xiangtan, Hunan, China, 411105.
	(\email{kaijiang@xtu.edu.cn}, \email{siwei@smail.xtu.edu.cn}). }
\and Wei Si\footnotemark[1]
\and Jie Xu\thanks{
	LSEC \& NCMIS, Institute of Computational Mathematics and
	Scientific/Engineering Computing (ICMSEC), Academy of Mathematics and
	Systems Science (AMSS), Chinese Academy of Sciences, Beijing, China
	(\email{xujie@lsec.cc.ac.cn}). } }

\usepackage{amsopn}

\makeatletter
\newcommand*{\addFileDependency}[1]{
  \typeout{(#1)}
  \@addtofilelist{#1}
  \IfFileExists{#1}{}{\typeout{No file #1.}}
}
\makeatother

\usepackage{url}
\usepackage{xcolor}
\definecolor{newcolor}{rgb}{.8,.349,.1}

\newcommand\tbbint{{-\mkern -16mu\int}}

\newcommand\dbbint{{-\mkern -19mu\int}}

\newcommand\bbint{
	{\mathchoice{\dbbint}{\tbbint}{\tbbint}{\tbbint}}
}

\DeclareMathOperator*{\argmin}{\mathrm{argmin}}
\DeclareMathOperator*{\argmax}{\mathrm{argmax}}

\newcommand{\mbk}{\bm{k}}

\newcommand{\mbr}{\bm{r}}

\newcommand{\mbt}{\bm{t}}

\newcommand{\mbbQ}{\mathbb{Q}}
\newcommand{\mbbR}{\mathbb{R}}
\newcommand{\mbbZ}{\mathbb{Z}}

\newcommand{\hphi}{\hat{\phi}}
\newcommand{\tphi}{\tilde{\phi}}
\newcommand{\hPhi}{\hat{\Phi}}
\newcommand{\tR}{\tilde{R}}
\newcommand{\tmbr}{\tilde{\mbr}}

\newcommand{\mcP}{\mathcal{P}}
\newcommand{\tmcP}{\tilde{\mcP}}
\newcommand{\msA}{\mathscr{A}}

\definecolor{SlateBlue1}{RGB}{131, 111, 255}
\definecolor{DarkOrange1}{RGB}{255, 127, 0}
\definecolor{DodgerBlue2}{RGB}{28, 134, 238}
\definecolor{DarkCyan}{RGB}{0, 139, 139}
\definecolor{DarkRed}{RGB}{139, 0, 0}
\definecolor{grey11}{RGB}{28, 28, 28}
\definecolor{Red1}{RGB}{255, 0, 0}
\definecolor{DarkMagenta}{RGB}{139, 0, 139}
\definecolor{lightGreen}{RGB}{144, 238, 144}

\ifpdf
\hypersetup{
  pdftitle={Tilt grain boundaries of hexagonal structures: a spectral viewpoint},
  pdfauthor={K. Jiang, W. Si, and J. Xu}
}
\fi

\definechangesauthor[name={Kai Jiang}, color=red]{JK}
\definechangesauthor[name={Wei Si}, color=violet]{SW}
\definechangesauthor[name={Xu Jie}, color=blue]{XJ} 

\begin{document}

\maketitle

\begin{abstract}
	We propose a spectral viewpoint for grain boundaries that are generally quasiperiodic. 
	To accurately capture the spectra computationally, it is crucial to adopt
	the projection method for quasiperiodic functions.
	Armed with the Lifshitz--Petrich free energy, we take the spectral viewpoint
	to examine tilt grain boundaries of the hexagonal phase.
	Several ingredients of grain boundaries are extracted, which are not easy to
	obtain from real-space profiles.
	We find that only a few spectra substantially contribute to the formation of
	grain boundaries.
	Their linear relation to the intrinsic spectra of the bulk hexagonal phase
	is independent of the tilt angle.
	By examining the feature of the spectral intensities, we propose a
	definition of the interface width.
	The widths calculated from this definition are consistent with visual estimation. 
\end{abstract}




\section{Introduction}
\label{sec:intro}

Grain boundaries (GBs) are transition regions between two grains of the same
ordered phase with different orientations.
Many experimental and theoretical studies have been performed on the GBs of
various systems, including metals\,\cite{buban2006grain, beyerlein2010statistical,
uberuaga2013point} and soft matters\,\cite{li2019sculpted, feng2021visualizing}.
GBs heavily affect material properties, such as strength,
plasticity\,\cite{ogata2005energy}, toughness\,\cite{watanabe2004toughening},
and corrosion resistance\,\cite{shimada2002optimization}. For this reason,
understanding GBs has long been a long central topic in material science.

Usually, GBs are regarded as consisting of several local imperfections compared
with the bulk structure.
Depending on how the local imperfections are deviated from the bulk structure,
they are classified into vacancies, interstitial atoms, or
dislocations\,\cite{brainBOOKgrain}.
Theories and algorithms are then developed to study the morphology, distribution
and dynamics of these local structures, including molecular dynamics
technique\,\cite{aguirre2019molecular, riet2021molecular}, Monte Carlo
simulation\,\cite{mason2015grain, son2020twodimensional}, continuum phase-field
approaches\,\cite{yamanaka2017phase, flint2019phasefield}, and multi-scale
methods\,\cite{madadi2021coarse}.
When studying GB structures, the relative orientation and displacement, as well
as the dividing planes of two grains, are essential variables.
Thus, when these variables are different the types of local structures appearing
would be distinct. 
If one focuses on the local structures, it seems difficult to take a unified viewpoint. 

In this paper, we propose a new spectral viewpoint to study the GB structures 
in a unified manner. 
When studying bulk structures, the spectral viewpoint and real-space
viewpoint are equally significant.
Only very few spectral points exhibit strong intensity, as evidenced by x-ray
diffractions for periodic crystals\,\cite{fischer2011colloidal} and
quasicrystals\,\cite{Shechtman1984}.
From the spectral points, we could extract the main constituents that
characterize the bulk structure.
In the mathematical language, we could effectively describe the bulk structure
by very few trigonometric functions.
It would be expected that a GB has similar property if we could appropriately
formulate its spectra, and that we could reveal some features distinct from the
real-space viewpoint.

Intuitively, a GB is formed approximately within a plane with certain direction.
Thus, the spectra of a GB shall be associated with the bulk profile restricted
on a plane with this direction.
This guides us to the concept of quasiperiodicity, since a quasiperiodic
function is defined by a restriction of a periodic function in a higher
dimension on a linear subspace.
Let us write down the definition of a $d$-dimensional quasiperiodic function. 
\begin{definition}\label{def:quasifun}
A function $f(\mbr)$ in $\mbbR^d$ is quasiperiodic if there exists a continuous
$d_0$-dimensional ($d_0\geq d$) periodic function $F$ which satisfies
$f(\mbr)=F(\mcP^T\mbr)$ for certain $d\times d_0$ matrix $\mcP$.
The matrix $\mcP$ is called the projection matrix. 
If we write out its columns as $\mcP=(\bm{p}_1, \cdots, \bm{p}_{d_0})$ with
$\bm{p}_i\in \mbbR^d$, they shall be linearly independent over $\mbbQ$.
\end{definition}
As a simple example, the function $f(x) = \sin(2\pi x) + \sin (2\pi \alpha x)$
is quasiperiodic for irrational $\alpha$, because we could choose $F=\sin(2\pi
x_1)+\sin(2\pi x_2)$ and $\mcP=(1,\alpha)$.
In other words, it is a restriction of $F=\sin(2\pi x_1)+\sin(2\pi x_2)$ on the
line $x_2=\alpha x_1$.
It is also noted that periodic functions are special quasiperiodic functions. 

Quasiperiodic functions have been investigated extensively since
$\mathrm{Poincar\acute{e}}$ studied dynamical systems with quasiperiodic
oscillations\,\cite{holmes1990poincare}.
The well-known approximation theorem denotes that any quasiperiodic function can
be approximated by a sequence of trigonometric
polynomials\,\cite{corduneanu1989almost}.
However, to efficiently approximate a quasiperiodic function, one may
necessarily include trigonometric functions with incommensurate periods. 
This point has not received proper attention in previous works, as we comment below.

For the approximation of a quasiperiodic function, an approach frequently
adopted is to use a periodic function with a delicately chosen period, called
the periodic approximation method (PAM).
The PAM has an unavoidable error, mainly from the Diophantine approximation, to
approximate irrational numbers by rational numbers.
To obtain an adequately precise quasiperiodic solution, the PAM requires a large 
unit cell to approximate the quasiperiodic system, oftentimes leading to
expensive and even unaccepted computational costs.
Furthermore, the rational approximation error of PAM does not decay uniformly as
the size of the computational domain increases.
The implementation and explanation of this method have been summarized in
\cite{jiang2014numerical}, and a rigorous mathematical theory can be found in
the forthcoming work\,\cite{jiang2021on}.
Another approach is the projection method (PM) based on
\Cref{def:quasifun}\,\cite{jiang2014numerical, jiang2018numerical}, to utilize a
higher-dimensional trigonometric polynomial.
By carefully choosing the projection matrix $\mcP$, many desired spectra can be
expressed precisely.
As a result, although discretizations in a higher dimension are involved, it
turns out that much fewer modes are needed so that a significantly lower
computational cost is reached.

The incorporation of quasiperiodicity into the computation of interfaces (note
that GBs are special interfaces) has emerged in a general
framework\,\cite{xu2017computing, cao2021computing}, and a few preliminary
results are obtained. 
In this framework, the whole space is divided into three regions by two parallel planes. 
The two bulk structures, with each one displaced or rotated, occupy the two on
the sides, so that the transition zone would occur in between.
The positions and orientations of the bulk structures are fixed by anchoring
boundary conditions on the two dividing planes. 
The corresponding function space is consistent with two chosen bulk structures.
This framework was first applied to computing interfaces between
three-dimensional periodic structures with matching period assumptions based on
experiments\,\cite{xu2017computing}. 
In a later work \cite{cao2021computing}, the computation framework has been
delicately designed again through employing the PM to handle quasiperiodicity
and Jacobi polynomials to deal with anchoring boundary conditions. 
A couple of interfacial structures between periodic structures and quasicrystals
have been obtained. 

Through this computational framework, one can obtain precise spectra along the
directions of two parallel dividing planes. 
However, the spectral information has not been examined previously. 
In this work, we investigate the spectra of tilt GBs of the hexagonal phase with
varying tilt angles, from which we reveal several features that cannot be easily
acquired via observing local morphology. 
We find that the morphology of GBs can be effectively represented by a few
primary spectral modes, and that GBs with different tilt angles have the
invariant representation about the spectral indices.
The intensities of the spectra show different features on oscillations far away
and close to the GB, from which we could define the interfacial width by a
formula about a few quantities characterizing the oscillations, rather than
through phenomenological observations. 
These results demonstrate the advantages of the spectral viewpoint, which is
prospective as a unified approach for the study of GBs between other bulk
phases. 

The rest of the paper is organized as follows. 
In \Cref{sec:funcspc}, we introduce the framework for computing GBs, 
especially the appropriate function space. 
The properties of the tilt GB system are also discussed. 
In \Cref{sec:setup}, we introduce the Lifshitz--Petrich (LP) free energy that we
utilize to study GBs, as well as the corresponding anchoring boundary
conditions. 
In \Cref{sec:discrete}, we present the spatial discretization for the LP free
energy and the optimization methods for the stationary states and optimal GBs. 
The spectra of tilt GBs between two hexagonal grains are discussed in \Cref{sec:results}. 
\Cref{sec:conclu} gives some concluding remarks.

\section{The function space of tilt GBs}
\label{sec:funcspc}

Since GBs are formed between two grains of the same bulk phase, we first need to
write down the bulk profile.
We assume that the bulk phase is described by a scalar field $\phi_0(\mbr)$
where $\mbr=(x,y,z)^T$ represents the spatial location.
We consider the general case that the profile $\phi_0(\mbr)$ is quasiperiodic.
By \Cref{def:quasifun}, it can be expressed based on a $d_0$-dimensional Fourier
series\,\cite{jiang2014numerical, jiang2018numerical},
\begin{equation}
  \phi_0(\mbr) = \sum_{\mbk\in\mbbZ^{d_{0}}} \hphi_0(\mbk) e^{i(\mcP\mbk)^{T}\cdot\mbr},
  \label{eq:bulk}
\end{equation}
where $i$ is the imaginary unit, and the projection matrix $\mcP\in
\mbbR^{3\times d_0}$ is rationally column full-rank. 
The integer vector $\mbk$ represents the indices of reciprocal lattice vectors.

To formulate a GB system, we need to specify the orientation and displacement of each grain, 
which can be described by two rotation matrices $R_{\pm}\in \rm{SO}(3)$ and two
vectors $\mbt_{\pm}\in\mbbR^{3}$, respectively.
Let us pose two grains with the dividing plane chosen as $x=0$. 
The subscript $-$ ($+$) represents the grain on the side $x<0$ ($x>0$). 
The profiles of two grains, denoted by $\phi_{\pm}(\mbr)$, are then given by 
\begin{equation}
	\begin{aligned}
		\phi_{\pm}(\mbr) &= \phi_{0}(R_{\pm}\mbr+\mbt_{\pm}) 
		= \sum_{\mbk\in\mbbZ^{d_{0}}} \hphi_{0}(\mbk)
			e^{ i(\mcP\mbk)^{T} \cdot (R_{\pm}\mbr+\mbt_{\pm}) } \\
		&= \sum_{\mbk\in\mbbZ^{d_{0}}} \left\{ \hphi_{0}(\mbk)
			e^{ i(\mcP\mbk)^{T}\cdot\mbt_{\pm} } \right\}
			e^{ i(R_{\pm}^{T}\mcP\mbk)^{T}\cdot\mbr } 
		= \sum_{\mbk\in\mbbZ^{d_{0}}} \hphi_{\pm}(\mbk)
			e^{ i(R_{\pm}^{T}\mcP\mbk)^{T}\cdot\mbr }, 
	\end{aligned}
	\label{eq:phi.Rt}
\end{equation}
where $\hphi_{\pm}(\mbk) = \hphi_{0}(\mbk) e^{i(\mcP\mbk)^{T}\cdot\mbt_{\pm}}$.
It is noticed that the rotation $R_{\pm}$ is imposed on the projection matrix
$\mcP$, while the displacement $\mbt_{\pm}$ acts on the Fourier coefficients.

\begin{figure}[!htbp]
	\centering
	\includegraphics[width=0.8\linewidth]{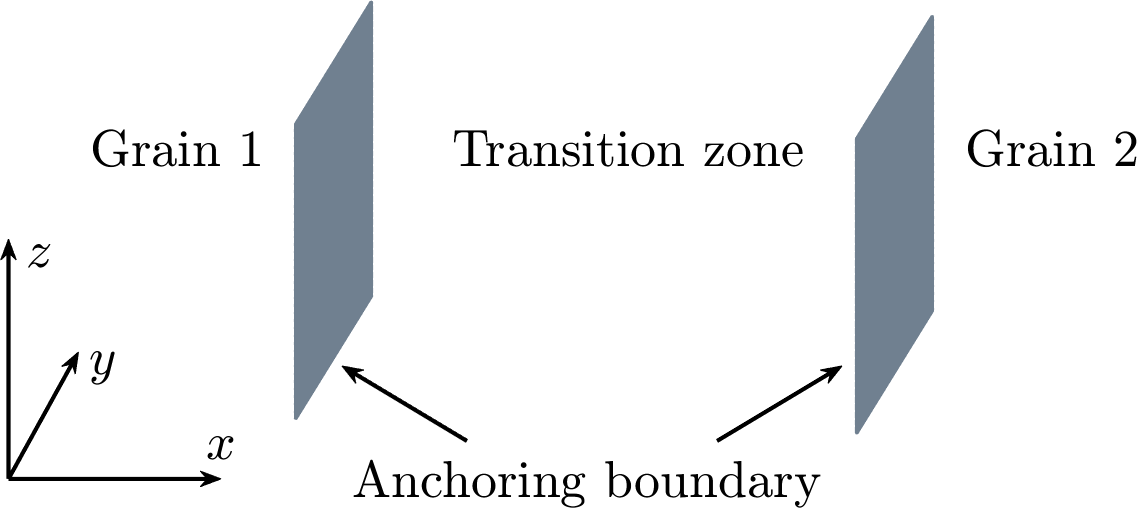}
	\caption{\label{fig:set}
		The setup of the interface system.
		}
\end{figure}

A GB is usually a relatively smooth transition between the grains. 
Thus, we divide the whole space into three regions by two parallel planes $x =
-L$ and $x = L$ for some $L$.
We assume that initially the two grains occupy the region $x \leq -L$ and $x
\geq L$, respectively, as shown in \cref{fig:set}.	
Hence, the GB is allowed to form in the region $-L < x < L$, which is our computational region.
The boundary conditions on the planes $x=\pm L$ rely on the model and are left to the next section. 
The function space along the $y$-$z$ plane depends on the rotation matrices
$R_{\pm}$ and the projection matrix $\mcP$, which we discuss below.

We shall look back into the rotated profiles $\phi_{\pm}(\mbr)$. 
Let us decompose the rotation matrix $R_{\pm}$ as $(R_{\pm x},\tR_{\pm})$, where
$R_{\pm x}$ is the first column, and $\tR_{\pm}$ is the second and third columns
of $R_{\pm}$.
Denote $\tmbr=(y,z)^T$. 
The phase profiles for two grains can be rewritten as 
\begin{equation}
	\phi_{\pm}(x,\tmbr) = \sum_{\mbk\in\mbbZ^{d_{0}}} 
		\left\{ \hphi_{\pm}(\mbk)
		e^{ i(R_{\pm x}^{T}\mcP\mbk) x } \right\}
		e^{ i(\tR_{\pm}^{T}\mcP\mbk)^{T}\cdot\tmbr }.
	\label{eq:phi.expan}
\end{equation}
It is clear that $\phi_{\pm}(x,\tmbr)$ lies within the function space 
\begin{equation}
	\msA_{\pm} = \left\{ \sum_{\mbk\in\mbbZ^{d_{0}}} a_{\pm}(x,\mbk)
		e^{ i(\tmcP_{\pm}\mbk)^{T}\cdot\tmbr } \right\},
	\label{eq:As.space}
\end{equation}
where $\tmcP_{\pm} = \tR_{\pm}^{T} \mcP$ is a $2\times d_{0}$ matrix. 
We can see that the two function spaces $\msA_{\pm}$ are also expressed as
projections of Fourier series from the $d_0$-dimensional space to a
two-dimensional space, with the projection matrix $\tmcP_{\pm}$ determined by
the bulk projection matrix $\mcP$ and the rotation matrices $R_{\pm}$.

To be consistent with both grains, it is then necessary to construct a suitable
function space that contains $\msA_{+}$ and $\msA_{-}$.
This can be done by extracting the linearly independent vectors from the
$2\times 2d_{0}$ matrix $(\tmcP_{+},\tmcP_{-})$, \textit{i.e.} identifying a $2\times d$
matrix $\tmcP$, column full-rank on $\mbbQ$, such that there exists a $d\times
2d_{0}$ integer matrix $Z$ satisfying
\begin{equation}
	\tmcP Z = (\tmcP_{+}, \tmcP_{-}).
\end{equation}
The function space is defined as
\begin{equation}
	\msA = \left\{ \phi(\mbr) = \sum_{\mbk\in\mbbZ^{d}} \hphi(x, \mbk)
		e^{ i(\tmcP\mbk)^{T} \cdot \tmbr } \right\}.
	\label{eq:A.space}
\end{equation}
It can be verified that $\msA_{\pm}\subseteq \msA$ from
\begin{equation}
	\tmcP_{+}\mbk = (\tmcP_{+},\tmcP_{-}) 
		\begin{pmatrix} \mbk \\ \bm{0} \end{pmatrix}
		= \tmcP Z \begin{pmatrix} \mbk \\ \bm{0} \end{pmatrix}.
\end{equation}
\eqref{eq:A.space} indicates that $\tmcP_{\pm}$ gives the spectra along the $y$-$z$ plane. 
Since $\tmcP_{\pm}$ is calculated from $\mcP$ and $R_{\pm}$, we can see that the
orientations of two grains affect the function space $\msA$. 

In this work, we would like to focus on tilt GBs with the bulk phase having a mirror plane. 
We could pose the bulk phase such that the mirror plane is exactly $x=0$, i.e. 
\begin{align}
	\phi_0(x,y,z) = \phi_0(-x,y,z).
	\label{eq:mirror}
\end{align}
Then, the rotations are done in a mirror-symmetric manner, \textit{i.e.} the
left grain is rotated by the angle $-\theta$ clockwise around the $z$-axis and the
right grain by the angle $\theta$.
In this case, the rotation matrix is $R(-\theta)$ for the left, and $R(\theta)$
for the right, where 
\begin{equation}
	R(\theta) = 
	\begin{pmatrix}
		\cos\theta & \sin\theta & 0 \\
		-\sin\theta & \cos\theta & 0 \\
		0 & 0 & 1
	\end{pmatrix}.
	\label{eq:rotate}
\end{equation}
The profiles for two grains are given by 
\begin{align*}
	\phi_{+}(x,y,z)=&~\phi_0(x\cos\theta+y\sin\theta,-x\sin\theta+y\cos\theta,z),\quad x>0;\\
	\phi_{-}(x,y,z)=&~\phi_0(x\cos\theta-y\sin\theta,x\sin\theta+y\cos\theta,z), \quad x<0. 
\end{align*}
Together with the symmetry of $\phi_0$, we can deduce that $\phi_{+}(x,y,z)=\phi_{-}(-x,y,z)$.

We turn to the spectra of $\phi_{+}$ and $\phi_{-}$.
From \cref{eq:mirror}, we have 
\begin{align}
	\hphi_0(\mbk) e^{i(\mcP\mbk)^{T}\cdot\mbr} = \hphi_0(\mbk) e^{i(D\mcP\mbk)^{T}\cdot\mbr},
	\quad D=\mathrm{diag}(-1,1,1). 
	\label{eq:indicate.mirror}
\end{align}
It implies that the columns in $D\mcP$ can be expressed linearly by the columns
in $\mcP$ by integer coefficients, \textit{i.e.} there is certain
$K\in\mbbZ^{d_0\times d_0}$ such that 
\begin{align}
	D\mcP=\mcP K. 
\end{align}
Therefore, we deduce that 
\begin{align}
  R^{T}(-\theta)\mcP = R^{T}(-\theta)D\mcP K = DR^{T}(\theta)\mcP K. 
\end{align}
Looking at the second and thrid components in the above, we actually arrive at 
\begin{align}
  \tmcP_{-}=\tR_{-}^T\mcP = \tR_{+}^{T}\mcP K=\tmcP_{+}K. 
\end{align}
Hence, when choosing the matrix $\tmcP$, we only need to focus on
$\tmcP_{-} = \tmcP_{+} K$, to find a column full-rank-in-$\mbbQ$ matrix $\tmcP$
such that $\tmcP Z' = \tmcP_{-}$ for some integer matrix $Z'$. 
In other words, for general GBs, we need to extract linearly independent vectors
from $2d_0$ vectors, while for tilt GBs, we only need to consider
$d_0$ vectors.

\begin{remark}
It is worth noting that the spectral viewpoint is embedded in the above formulation. 
For a bulk structure, its spectra can be seen clearly only when the profile is
written in the form \cref{eq:bulk}, where the $\mcP$ matrix is predetermined.
For a GB, its spectra are formulated in the $y$-$z$ plane as \cref{eq:A.space} presents.
\end{remark}

The construction of the function space $\msA$ requires considerable efforts. 
One might be tempted to propose some simple functions spaces, such as a large
cell with simple (periodic or Neumann) boundary conditions.
It is not difficult to find that except under special orientations, the boundary
conditions cannot fit the bulk structure of two grains. 
How big computational cell is enough to accommodate the incommensurability or 
quasiperiodicity? 
In fact it corresponds to a well-known problem which deals with the
approximation of real numbers by rational numbers, called the
Simultaneous Diophantine Approximation in number theory.
The approximation error cannot be eliminated no matter how large the
computational cell is\,\cite{jiang2014numerical, jiang2021on}.
From the spectral viewpoint, large cell computations are inconsistent with the
structure of the spectra which we believe is essential in the GB system.

\section{Free energy}
\label{sec:setup}

We introduce the Lifschitz-Petrich model and write down the free energy for the
bulk phases and the GB systems.

\subsection{Lifshitz--Petrich free energy functional}

Inspired by quasiperiodic Faraday wave
experiments\,\cite{edwards1993parametrically}, Lifshitz and Petrich imposed two
characteristic length scales into the Landau free energy functional to stabilize
two frequencies in ordered structures, which might be periodic or
quasiperiodic\,\cite{lifshitz1997theoretical, jiang2015stability}.
With a rescaling, the Lifshitz--Petrich (LP) free energy density about a scalar
order parameter $\phi$ is given by 
\begin{equation}
	E[\phi(\mbr);\Omega] = \frac{1}{V(\Omega)} \int_{\Omega} \left\{ 
		\frac{1}{2} [(\Delta+1)(\Delta+q^{2})\phi]^{2}
		-\frac{\epsilon}{2}\phi^{2} - \frac{\alpha}{3}\phi^{3} 
			+ \frac{1}{4}\phi^{4} \right\} d\mbr,
	\label{eq:lp.model}
\end{equation}
where $\Omega$ is a region in $\mbbR^{3}$, whose volume is $V(\Omega)$. 
As for the parameters, $q$ is the ratio of two critical wave lengths, 
$\epsilon$ can be comprehended as the reduced temperature, and $\alpha$ is a
phenomenological parameter.
Associated with the free energy is the mass conservation, 
\begin{equation}
	\frac{1}{V(\Omega)} \int_{\Omega} \phi(\mbr) d\mbr = \bar{\phi},
	\label{eq:lp.mass}
\end{equation}
where the value of the constant $\bar{\phi}$ could vary in different cases,
which we will specify.

\subsection{LP model for bulk profiles}

The bulk profile of a phase is obtained by minimizing the functional
\cref{eq:lp.model} when taking the limit $\Omega\to\mbbR^{3}$ and letting
$\bar{\phi}$ tend to zero under this limit. 
If the phase is periodic with the unit cell $\Omega_{0}$, we can verify that
\begin{equation}
	\lim_{\Omega\to\mbbR^{3}} E[\phi(\mbr);\Omega] = E[\phi(\mbr);\Omega_{0}],
	\label{eq:box.limit}
\end{equation}
which is the energy density in the unit cell.
On the other hand, the limit on the left-hand side is also suitable for
quasicrystals.
For periodic or quasiperiodic phases, taking the profile \cref{eq:bulk} into
the free energy \cref{eq:lp.model} and noticing 
\begin{equation}
	\lim_{\Omega\to\mbbR^{3}} \frac{1}{V(\Omega)} \int_{\Omega}
	e^{i(\mcP\mbk)^{T}\cdot\mbr} d\mbr = \delta(\mcP\mbk),
	\label{eq:dirac}
\end{equation}
where $\delta(\cdot)$ is the Dirac function, we obtain
\begin{equation}
	\begin{aligned}
		\lim_{\Omega\to\mbbR^{3}} &E[\phi(\mbr);\Omega] 
		= \frac{1}{2} \sum_{\mbk\in\mbbZ^{d_{0}}} \left( (1-|\mcP\mbk|^{2})^{2}
		(q^{2}-|\mcP\mbk|^{2})^{2} - \epsilon \right) \hphi(\mbk) \hphi(-\mbk)
		\\
		&- \frac{\alpha}{3} \sum_{\mbk_{1}+\mbk_{2}+\mbk_{3}=\bm{0}} 
			\hphi(\mbk_{1}) \hphi(\mbk_{2}) \hphi(\mbk_{3})
		+ \frac{1}{4} \sum_{\mbk_{1}+\mbk_{2}+\mbk_{3}+\mbk_{4}=\bm{0}} 
			\hphi(\mbk_{1}) \hphi(\mbk_{2}) \hphi(\mbk_{3}) \hphi(\mbk_{4}).
	\end{aligned}
	\label{eq:lp.model.pm}
\end{equation}
With an appropriate choice of $\mcP$, one could search the minimizer of
\eqref{eq:lp.model.pm} to obtain the bulk profile $\phi_{0}(\mbr)$. 
By imposing the orientation and displacement on $\phi_{0}(\mbr)$, the bulk
phases $\phi_{\pm}(\mbr)$ can be obtained by \eqref{eq:phi.expan}.

\subsection{LP model for GBs}

The computation of GBs is carried out in the banded region $-L<x<L$. 
In the $y$-$z$ plane, the GB structure is space-filling.
Thus, we define the average spatial integral over the $y$-$z$ plane as 
\begin{equation}
	\bbint = \lim_{R\to\infty} \frac{1}{V(C_{R})} \int_{C_{R}},
	\label{eq:bbint}
\end{equation}
where $C_{R} \subseteq \mbbR^{2}$ is a circle centered at the origin with radius
$R$ in the $y$-$z$ plane.
Hence, when computing GBs, the free energy \cref{eq:lp.model} can be rewritten as
\begin{equation}
	\begin{aligned}
	E(\phi) = &\underbrace{\frac{1}{2L} \int_{-L}^{L} \bbint \left\{ 
		\frac{1}{2} [(\Delta+1)(\Delta+q^{2})\phi]^{2} \right\} d\tmbr dx}_{G(\phi)} \\
		&+ \underbrace{\frac{1}{2L} \int_{-L}^{L} \bbint \left\{
			-\frac{\epsilon}{2}\phi^{2} - \frac{\alpha}{3}\phi^{3} 
			+ \frac{1}{4}\phi^{4} \right\} d\tmbr dx}_{F(\phi)},
	\end{aligned}
	\label{eq:lp.model.frame}
\end{equation}
where we recall $\tmbr = (y,z)^{T}$. 
For the convenience of the discussions afterwards, we divide the energy into two
parts: $G(\phi)$ is the interaction energy that contains high-order differential
operators to form ordered structures, and $F(\phi)$ is the bulk energy with the
polynomial-type formulation.
The mass constraint is
\begin{equation}
	\frac{1}{2L} \int_{-L}^{L} \bbint \phi(\mbr) d\tmbr dx = \frac{1}{2L}
	\left(\int_{-L}^{0} \bbint \phi_{-}(\mbr) d\tmbr dx
        +\int_{0}^{L} \bbint \phi_{+}(\mbr) d\tmbr dx\right).
	\label{eq:lp.mass.frame}
\end{equation}
Moreover, the Dirichlet boundary conditions are given on the function values and
normal derivatives of $\phi$ up to third order.
They shall be identical to the bulk values, \textit{i.e.},
\begin{equation}
	\left.\frac{\partial^{m} \phi(x,\tmbr)}{\partial x^{m}} \right|_{x=-L}
	= \left.\frac{\partial^{m} \phi_{-}(x,\tmbr)}{\partial x^{m}} \right|_{x=-L},
	~~
	\left.\frac{\partial^{m} \phi(x,\tmbr)}{\partial x^{m}} \right|_{x=L}
	= \left.\frac{\partial^{m} \phi_{+}(x,\tmbr)}{\partial x^{m}} \right|_{x=L},
	\label{eq:lp.bd.frame}
\end{equation}
where $m = 0,1,2,3$.

\section{Numerical details}
\label{sec:discrete}

We discretize the LP free energy \cref{eq:lp.model.frame} in
space by a spectral method with a combination of the projection method and
spectral-Galerkin method using generalized Jacobi polynomials.


We need to approximate the function space $\msA$ given by \cref{eq:A.space} by a
finite-dimensional space. 
In the $y$-$z$ direction, the function has already been expanded in the Fourier series.
Thus, we just choose an integer $N_F$ and truncate $\mbk=(k_1,\cdots,k_d)$ at
$|k_l|\le N_F$ to obtain 
\begin{equation}
	\phi(x,\tmbr) \approx \sum_{k_{l}\le N_F} \hphi(x, \mbk) 
		e^{i (\tmcP\mbk)^{T}\cdot\tmbr},
		~~~~ \tmbr = (y,z)^{T}.
	\label{eq:phi.expansion}
\end{equation}

In the $x$-direction, we approximate $\hphi(x,\mbk)$ using Jacobi polynomials.
For the inhomogeneous boundary conditions \cref{eq:lp.bd.frame}, we can
construct a seventh-order polynomial. 
It remains to consider the approximation function space for homogeneous boundary conditions,
\begin{equation}
	W_{N} = \mbox{span}\left\{ \varphi\in P_{N}: \left.
		\frac{\partial^{m}\varphi(x)}{\partial x^{m}} \right|_{x=-L}
		 = \left.\frac{\partial^{m}\varphi(x)}{\partial x^{m}} \right|_{x=L} = 0, 
		~~ m = 0,1,2,3 \right\}.
	\label{eq:lp.bd.homo}
\end{equation}
By the generalized Jacobi polynomial $J_{k}^{-4,-4}(x)$, a set of basis
functions of $W_{N}$ can be given by 
\begin{equation}
	\varphi_{l}(x) := J_{l+7}^{-4,-4}(x), ~~ l=0,\cdots,N-8.
\end{equation}
Thus, we could write 
\begin{equation}
	\hphi(x,\mbk) = \sum_{j=0}^{N-8} \tphi_{j}(\mbk) \varphi_{j}(x)
	= \sum_{j=0}^{N-8} \tphi_{j}(\mbk) J_{j+7}^{-4,-4}(x).
\end{equation}
More details can be found in \cite{cao2021computing}.

The integral in \eqref{eq:lp.model.frame} can be calculated accurately.
We notice that the integrand of $F(\phi)$ is a fourth-order polynomial.
Thus, we focus on calculating the highest-order term, and can handle two
lower-order terms in the same way.
Since $u$ is a polynomial of degree less than or equal to $N$, we find that the
fourth-order polynomial of degree less than or equal to $4N$ can be accurately
integrated if we use the Legendre Gauss quadrature of degree $2N$, \textit{i.e.},
\begin{equation}
	\frac{1}{2L} \int_{-L}^{L} \bbint u^{4}(x, \tmbr) d\tmbr dx 
	= \frac{1}{2L} \sum_{j=1}^{2N} \omega_{j}
		\sum_{\mbk_{1}+\mbk_{2}+\mbk_{3}+\mbk_{4}=\mathbf{0}} 
		\hat{u}(x_{j},\mbk_{1}) \hat{u}(x_{j},\mbk_{2}) 
		\hat{u}(x_{j},\mbk_{3}) \hat{u}(x_{j},\mbk_{4}),
	\label{eq:define.inner.frame}
\end{equation}
where $(x_{j},\omega_{j})$ are the Legendre Gauss points and weights.
Note that the summation about $\mbk$ can be computed by using FFT in $O\left(
(N_{F}\log N_{F})^{d} \right)$ operations.

Together with \cref{eq:phi.expansion} and \cref{eq:define.inner.frame}, the LP
free energy functional \cref{eq:lp.model.frame} is discretized as
\begin{equation}
	E_{\mbk}(\hPhi) = G_{\mbk}(\hPhi) + F_{\mbk}(\hPhi),
	\label{eq:lp.model.apg}
\end{equation}
where $G_{\mbk}$ and $F_{\mbk}$ are the discretized interaction and bulk energies,
\begin{equation}
	\begin{aligned}
		G_{\mbk}(\hPhi) &= \frac{1}{4L} \sum_{|\mbk|\leq N_{F}}
			\sum_{j=1}^{2N} \omega_{j}
			(1-|\tmcP\mbk|^{2}+\partial_{x}^{2})^{2} 
			(q^{2}-|\tmcP\mbk|^{2}+\partial_{x}^{2})^{2}
			\hphi(x_{j},\mbk) \hphi(x_{j},-\mbk),
	\end{aligned}
	\label{eq:lp.model.apg.g}
\end{equation}
\begin{equation}
	\begin{aligned}
		F_{\mbk}(\hPhi) &= \frac{1}{2L} \sum_{j=1}^{2N} \omega_{j} \left\{
			- \frac{\epsilon}{2} \sum_{|\mbk|\leq N_{F}}
				\hphi(x_{j},\mbk) \hphi(x_{j},-\mbk)\right.\\
		&~~~~~~~~~~~~~~~~~~~~
			- \frac{\alpha}{3} \sum_{\mbk_{1}+\mbk_{2}+\mbk_{3}=\bm{0}} 
				\hphi(x_{j},\mbk_{1}) \hphi(x_{j},\mbk_{2}) 
				\hphi(x_{j},\mbk_{3})  \\
		&~~~~~~~~~~~~~~~~~~~~
		\left.+ \frac{1}{4} \sum_{\mbk_{1}+\mbk_{2}+\mbk_{3}+\mbk_{4}=\bm{0}}
			\hphi(x_{j},\mbk_{1}) \hphi(x_{j},\mbk_{2}) 
			\hphi(x_{j},\mbk_{3}) \hphi(x_{j},\mbk_{4}) \right\}.
	\end{aligned}
	\label{eq:lp.model.apg.f}
\end{equation}
$|\mbk_{s}|\leq N_{F}$, $s=1,2,3,4$, and
\begin{equation*}
	\hPhi = (\hphi(x_{1},\mbk_{1}), \cdots, \hphi(x_{1},\mbk_{N_{F}}), \cdots,
	\hphi(x_{2N},\mbk_{1}), \cdots, \hphi(x_{2N},\mbk_{N_{F}})) \in
	\mathbb{C}^{2NN_{F}}.
\end{equation*}

It is evident that the nonlinear terms in $F_{\mbk}$ are $d$-dimensional
convolution in the reciprocal space.
A direct evaluation of these convolution terms is extremely expensive.
Instead, these terms are simple multiplication in the $d$-dimensional physical
space.
Similar to the pseudospectral approach, these convolutions can be efficiently
calculated through the FFT.
Moreover, the mass conservation constraint \cref{eq:lp.mass} is discretized as
\begin{equation}
	e_{\omega}^{T} \hPhi = 0, ~
	e_{\omega} = ( \underbrace{\omega_{1},0,\cdots,0}_{N_{F}}, \cdots,
		\underbrace{0,\cdots,\omega_{j},\cdots,0}_{N_{F}}, \cdots,
		\underbrace{0,\cdots,0,\omega_{2N}}_{N_{F}} )^{T} \in \mbbR^{2NN_{F}}.
	\label{eq:lp.mass.apg}
\end{equation}
Therefore, we obtain the following finite-dimensional minimization problem
\begin{equation}
	\min_{\hPhi\in\mathbb{C}^{2NN_{F}}} E_{\mbk}(\hPhi) 
	= G_{\mbk}(\hPhi) + F_{\mbk}(\hPhi),
	~~~~ \textit{s.t.} ~~ e_{\omega}^{T}\hPhi = 0.
	\label{eq:minization}
\end{equation}


Solving such a minimization problem requires an appropriate initial state.
Since the two bulk phases in the interface system are consistent in function
space $\msA$ \cref{eq:A.space}, the initial state can be constructed by connecting
them in a simple way,
\begin{equation}
	\hphi(x,\mbk) = (1-b(x))\hphi_{-}(x,\mbk) + b(x)\hphi_{+}(x,\mbk),
	\label{eq:connect}
\end{equation}
where $b(x)$ is a smooth monotone function satisfying $b(-L) = 0$ and $b(L) = 1$.
A good approximation to $b(x)$ is
\begin{equation}
	b(x) = \frac{1 - \tanh(\sigma x)}{2},
	\label{eq:bx}
\end{equation}
with $\sigma$ large.
Thus initial condition is
\begin{equation}
	\phi(x,\tmbr,0) = \sum_{|\mbk|\leq N_{F}} \left[ (1-b(x)) \phi_{+}(x, \mbk)
		+ b(x) \phi_{-}(x, \mbk) \right] e^{ i(\tmcP\mbk)^{T}\cdot\tmbr }.
	\label{eq:phi.init}
\end{equation}

The minimization problem \eqref{eq:minization} with the initial condition
\eqref{eq:phi.init} can be efficiently solved by the AA-BPG approach.
The approach combines the Nesterov acceleration technique, the backtracking
linear search method, and the restart strategy to reach the steady state efficiently.
In addition, this approach overcomes the requirement of global Lipschitz constant
and theoretically proves its convergence.
We refer to \cite{jiang2020efficient} for implementation details.

\section{GBs between hexagonal crystals}
\label{sec:results}

In this section, we examine the GBs of the hexagonal phase with varying tilt
angles, in particular focus on its spectra in the $y$-$z$ plane.

To begin with, we introduce the bulk profile and spectra of the hexagonal phase.
The hexagonal phase is periodic in two directions and homogeneous in the third.
Let us pose the homogeneous direction as $z$. 
In this case, the matrix $\mcP$ in \cref{eq:bulk} can be chosen as
\begin{equation}
	\mcP = \begin{pmatrix} 1 & \frac{1}{2} \\ 0 & \frac{\sqrt{3}}{2} 
		\\ 0 & 0 \end{pmatrix}.
	\label{eq:hex.mcP}
\end{equation}
The two columns of \cref{eq:hex.mcP} are actually two basic vectors that give
the parallelogram unit cell in the spectral space. 

\begin{figure}[!htbp]
	\centering
	\includegraphics[width=0.55\linewidth]{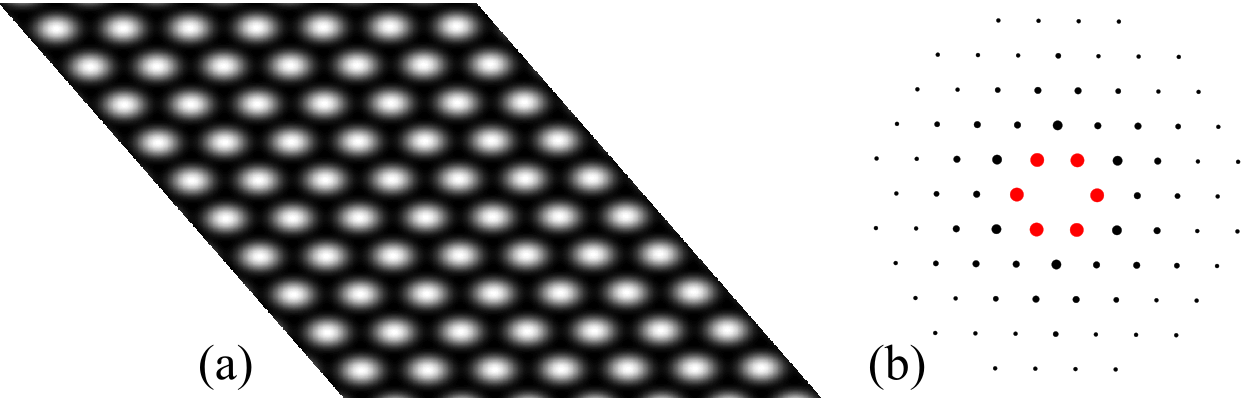}
	\caption{\label{fig:hex}
		Hexagonal crystal: (a) The real-space morphology; (b) spectra modes
		whose intensities are greater than $10^{-6}$.
		The size of these points reflects their intensity and the six strongest
		points marked by red.
		}
\end{figure}

By minimizing \cref{eq:lp.model.pm}, we obtain the bulk profile $\phi_0$. 
Its real-space morphology is presented in \cref{fig:hex} (a) with a few unit cells. 
The corresponding spectral distribution is given in \cref{fig:hex} (b). 
We label the six spectra with the strongest intensity (the intensity of a
spectrum is the absolute value of its coefficient) in red.
They play a vital role in the formation of the hexagonal structure. 
It turns out that the intensities of these six spectra are at least ten times
greater others.

\begin{figure}[!htbp]
	\centering
	\includegraphics[width=0.7\linewidth]{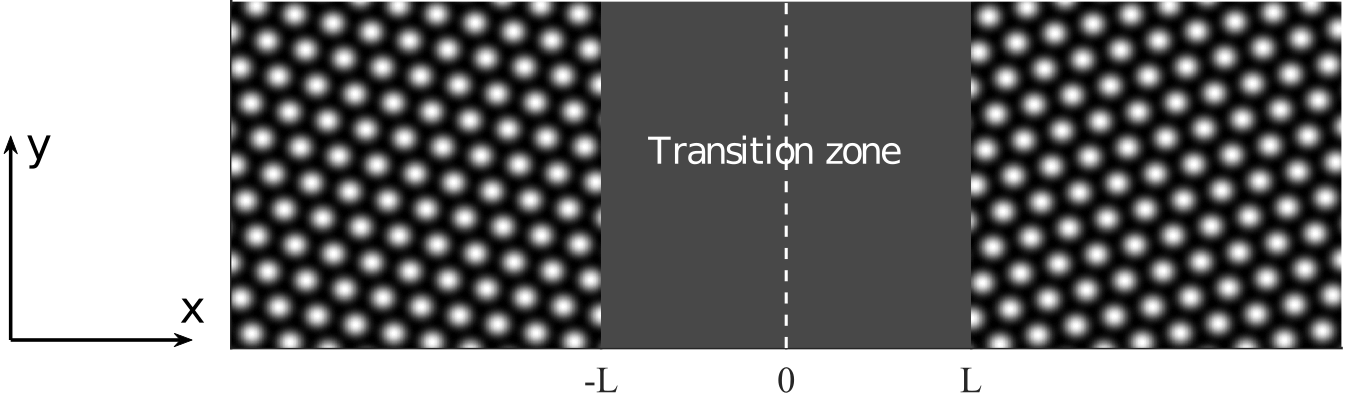}
	\caption{\label{fig:set.hex}
		The GB of hexagonal crystals in the $x$-$y$ plane.
		The left (right) grain in \cref{fig:hex} is rotated by $-\theta$
		($\theta$) clockwise around the $z$-axis.
		The transition region is the middle dark part between two anchoring
		planes at $x = \pm L$.
		The white dotted line is the mirror plane $x = 0$.
		}
\end{figure}

In the GB system, we still pose the grains such that $z$ is the homogeneous direction.
From the observation of the $x$-$y$ plane, we give a schematic pattern of the GB
system as shown in \cref{fig:set.hex}.
The grain on the left is rotated from the bulk profile (see \cref{fig:hex} by
the angle $-\theta$ clockwise around the $z$-axis, and the right one by $\theta$.
Two anchoring planes are located on $x = -L$ and $x = L$ and the mirror plane is $x = 0$.
The GB structure is computed in the middle dark part between the two planes. 

Next, we shall write down the matrix $\tmcP_{\pm}$.
The discussion in \cref{sec:funcspc} demonstrates that $\tmcP_{-} = \tmcP_{+}K$
for some integer matrix $K$, so that we only need to find a rationally column
full-rank matrix $\tmcP$ satisfying $\tmcP Z' =\tmcP_{+}$ for some integer
matrix $Z'$.
From \cref{eq:rotate} and \cref{eq:hex.mcP}, the projection matrix of the
right grain is
\begin{equation}
	\tmcP_{+} = \tR_{+} \mcP = \begin{pmatrix} 
		\sin\theta & \frac{1}{2}\sin\theta + \frac{\sqrt{3}}{2}\cos\theta
		\\ 0 & 0 \end{pmatrix}.
\end{equation}
We shall consider the general case where $\tan\theta/\sqrt{3}$ is not a rational
number, so the column rank of $\tmcP$ is $2$ over $\mbbQ$.
In this case, the system is quasiperiodic in the $y$-direction. 
The matrix $\tmcP$ can be chosen as 
\begin{equation}
	\tmcP = \begin{pmatrix} \frac{1}{2}\sin\theta & \frac{\sqrt{3}}{2}\cos\theta 
		\\ 0 & 0 \end{pmatrix}.
\end{equation}
With such a common projection matrix $\tmcP$, the actual spectral modes are given by
\begin{equation}
	\tmcP\mbk = \begin{pmatrix} \frac{k_{1}}{2}\sin\theta +
	\frac{\sqrt{3}k_{2}}{2}\cos\theta \\ 0 \end{pmatrix},
	\label{eq:actual.spectra}
\end{equation}
whose spectral index is $\mbk = (k_{1}, k_{2})^{T}\in\mbbZ^2$. 
The zero element in \cref{eq:actual.spectra} is generated from the homogeneous
$z$-direction in the GB system.
Therefore, a spectral point shall be regarded as a scalar, \textit{i.e.}, the
first element of \cref{eq:actual.spectra}.
With different tilt angles $\theta$, the scalar value varies. 

In the following simulations, the parameters in the LP free energy are chosen as $q =
2\cos(\pi/12)$, $\epsilon = 0.04$, $\alpha = 1$.
The size of spatial discretization in the $x$-direction is fixed at $N = 256$,
and $2N = 512$ Gauss points are used in the numerical integration.
For the length of the computational domain in the $x$-direction, we choose $L = 40\pi$. 
The Fourier series in the $y$-$z$ plane is truncated at $N_{F} = 20$.
Such a setting turns out to be sufficient, as the increase in discretization
points does not affect the results.

\begin{figure}[!htbp]
	\centering
	\subfigure[Tilt GB with $11\pi/63$]{\label{fig:GB.realwidth.11}
		\includegraphics[width=0.48\linewidth]{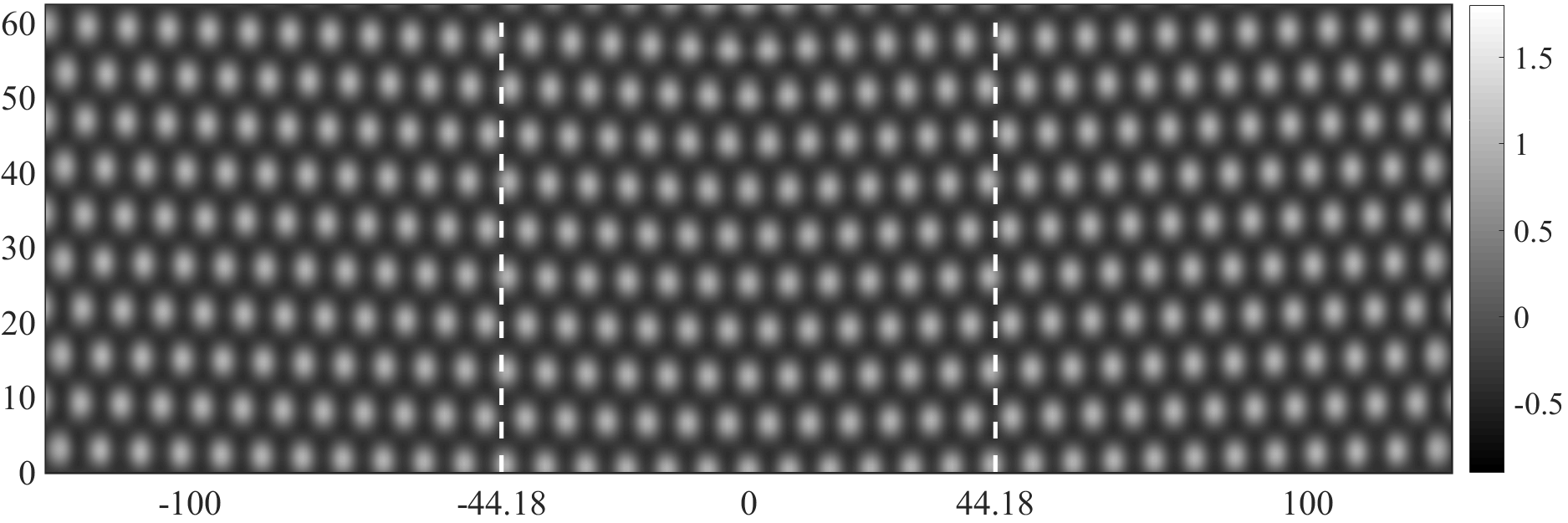}}
	\subfigure[Tilt GB with $12\pi/63$]{\label{fig:GB.realwidth.12}
		\includegraphics[width=0.48\linewidth]{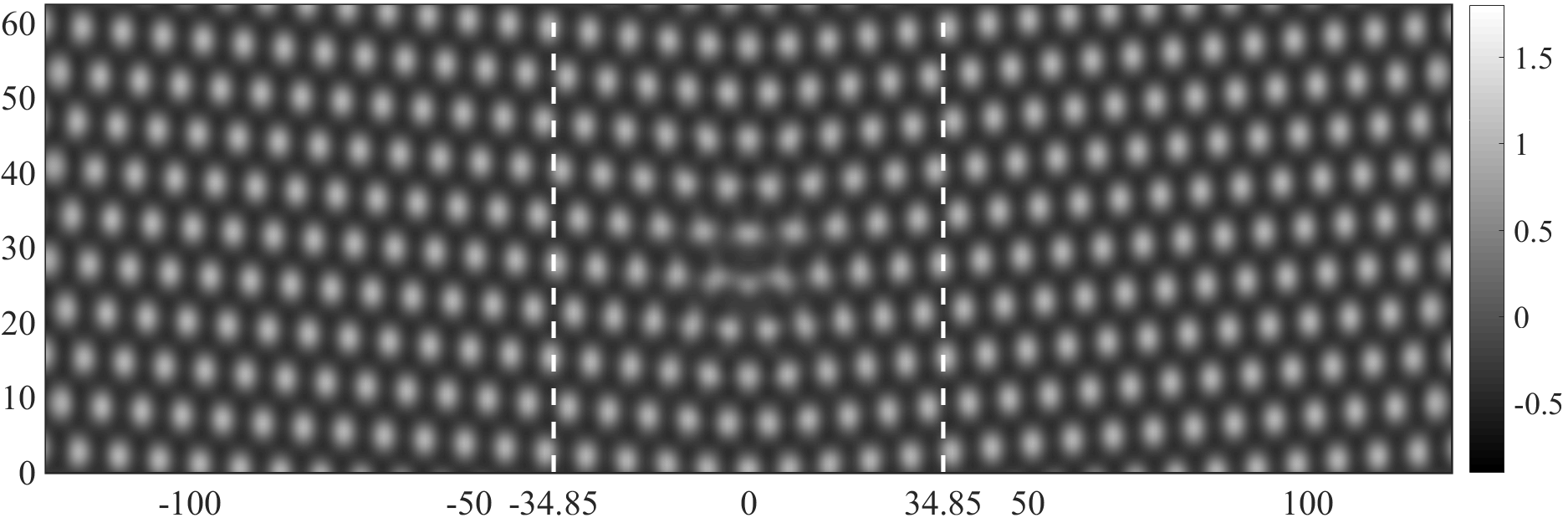}}
	\subfigure[Tilt GB with $13\pi/63$]{\label{fig:GB.realwidth.13}
		\includegraphics[width=0.48\linewidth]{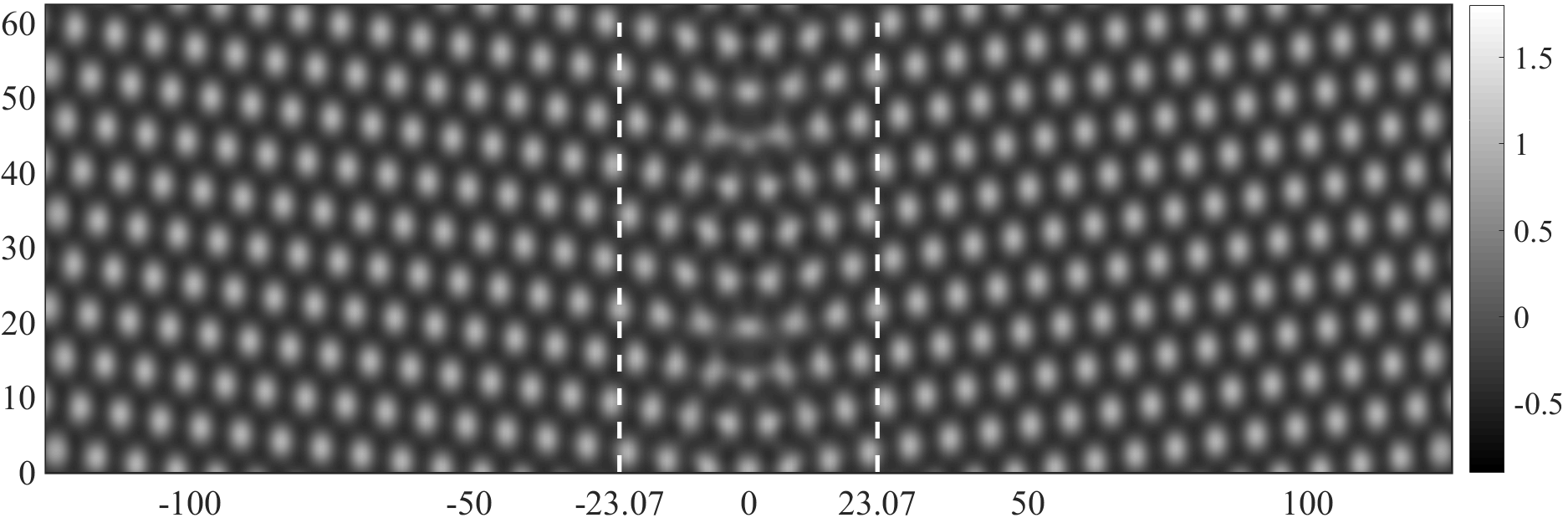}}
	\subfigure[Tilt GB with $14\pi/63$]{\label{fig:GB.realwidth.14}
		\includegraphics[width=0.48\linewidth]{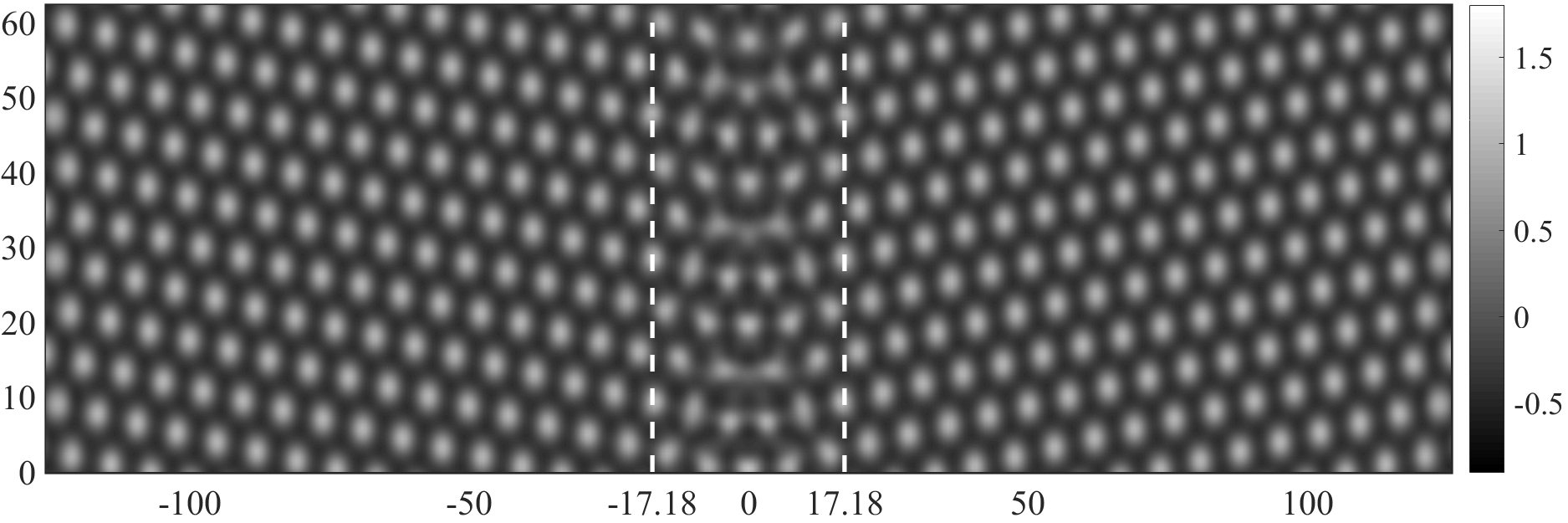}}
	\subfigure[Tilt GB with $15\pi/63$]{\label{fig:GB.realwidth.15}
		\includegraphics[width=0.48\linewidth]{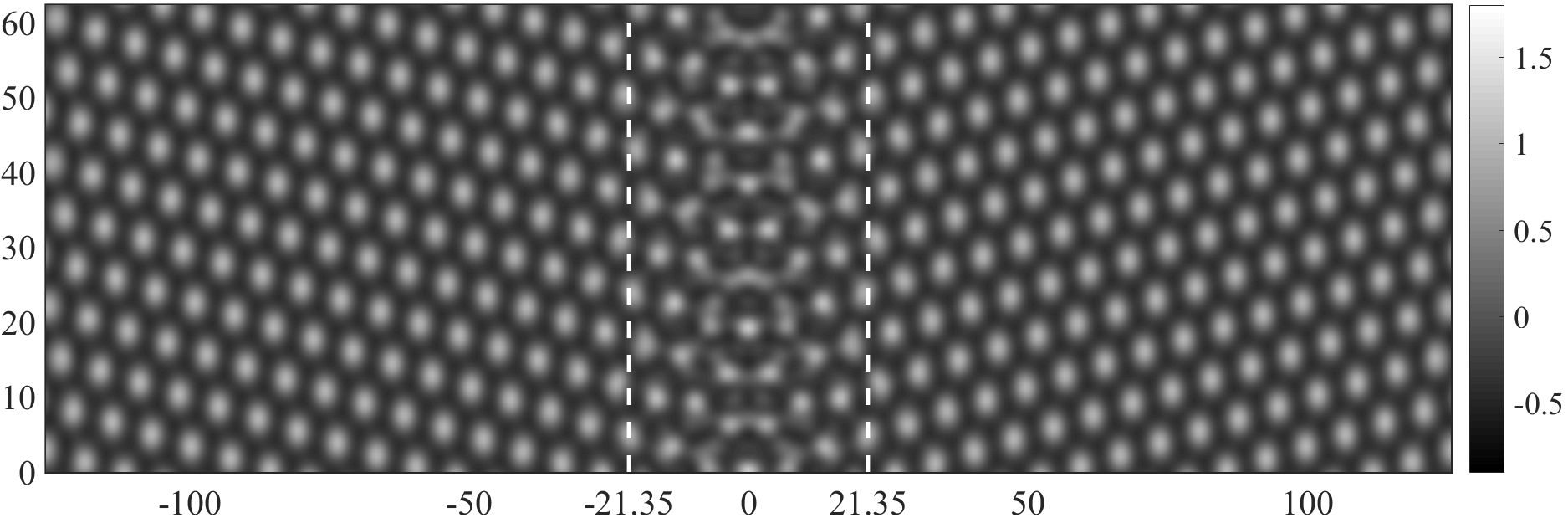}}
	\subfigure[Tilt GB with $17\pi/63$]{\label{fig:GB.realwidth.17}
		\includegraphics[width=0.48\linewidth]{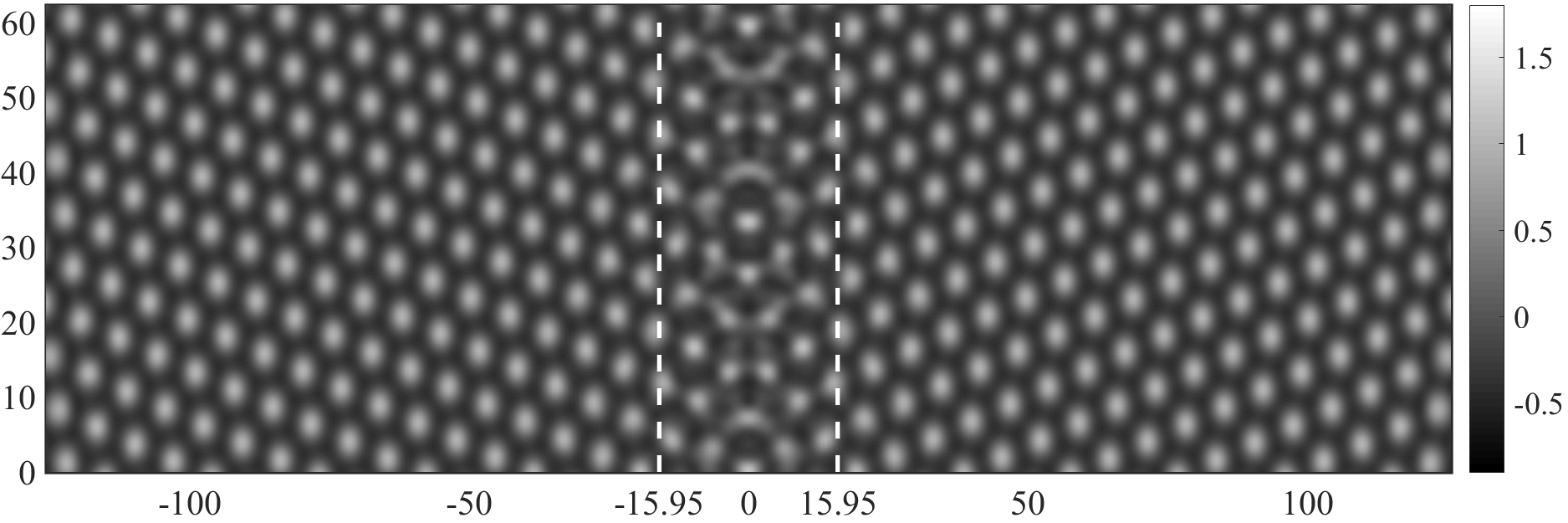}}
	\subfigure[Tilt GB with $18\pi/63$]{\label{fig:GB.realwidth.18}
		\includegraphics[width=0.48\linewidth]{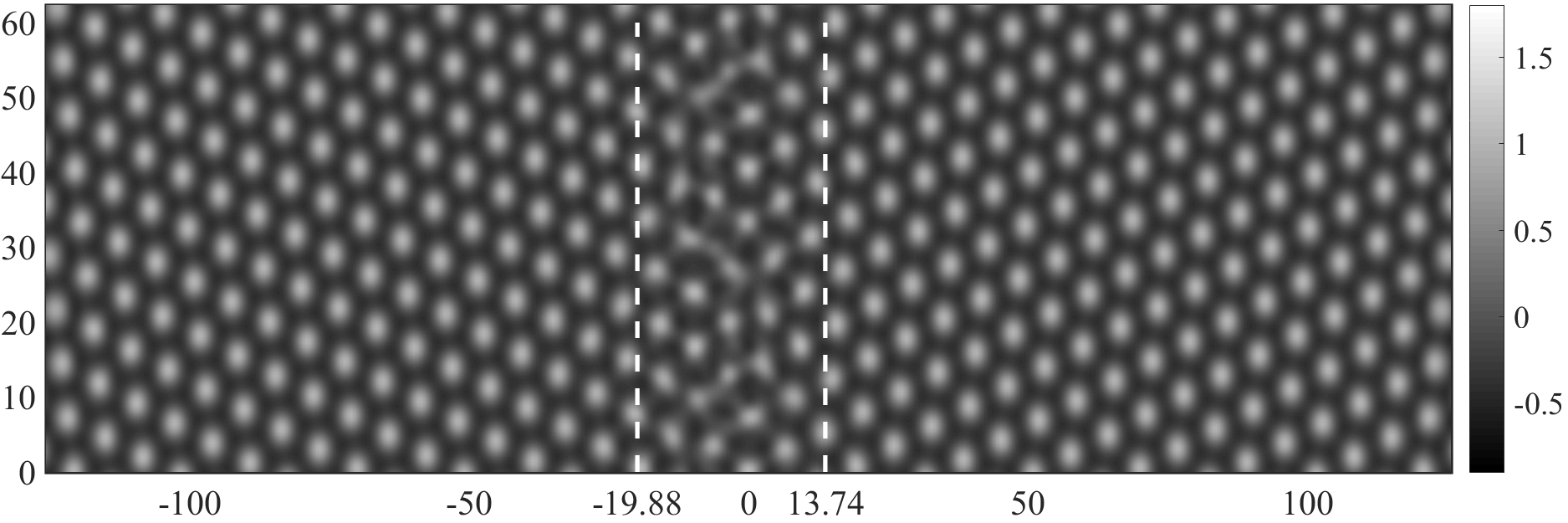}}
	\subfigure[Tilt GB with $19\pi/63$]{\label{fig:GB.realwidth.19}
		\includegraphics[width=0.48\linewidth]{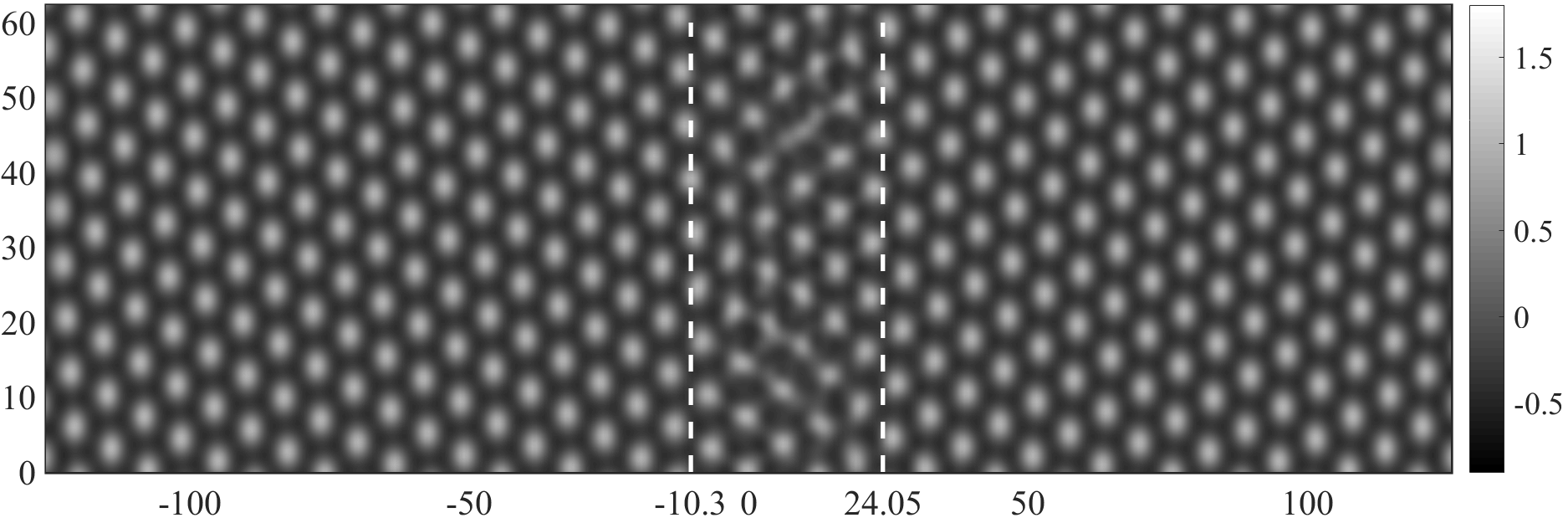}}
	\caption{\label{fig:GB.realwidth.cases}
		The real-space morphology of the hexagonal GB with various tilt angles.
		The interface edge is marked by two white dotted lines.
	}
\end{figure}

We have done simulations for many tilt angles to understand the mechanism of tilt GBs. 
Here, as representative cases, we only present the results for $\theta=11\pi/63,
12\pi/63, \cdots, 19\pi/63$, and discuss them from the spectral viewpoint.
The corresponding real-space morphologies are shown in
\cref{fig:GB.realwidth.cases}, where we do not include the morphology of
$\theta=16\pi/63$ as it will be discussed in detail afterwards.
We also label the interface width in the figure, which will be discussed in \Cref{subsec:width}. 
It should be emphasized that the GB is an infinite structure along the $y$-direction. 
Our computational framework can obtain the infinite quasiperiodic phase. 
In this work, we only show a part of the GB structure of $0\leq y \leq 20\pi$.

\subsection{Primary spectral modes}
\label{subsec:distribution.intensity}

It is seen in \cref{fig:hex} that in the bulk profile only six spectra dominate. 
For the GBs, we would like to examine whether this statement still holds. 
By choosing a threshold, we could classify the spectra into primary and
non-primary spectral modes by the intensities of Fourier coefficients
$|\hphi(x,\mbk)|$ for fixed $x$ and $\mbk$.
To be specific, we define the \textbf{primary spectral indices (PSI)} on certain
$x$-slice as
\begin{equation}
	\Lambda_{p}(x) = \left\{ \mbk: |\hphi(x,\mbk)| > \varepsilon \right\},
	~~~~ \forall x\in\mbbR,
	\label{eq:PSI.def}
\end{equation}
where $\hphi(x,\mbk)$ describes the spectra of the GB profile, and 
$\varepsilon$ is a threshold value.
Note that $\mbk\in\Lambda_{p}(x)$ implies that $-\mbk\in\Lambda_{p}(x)$ because
$\hphi(x,\mbk)$ and $\hphi(x,-\mbk)$ are complex conjugates by the fact that
$\phi$ is real-valued. 
It is crucial to choose an appropriate threshold value $\varepsilon$, which we discuss below.

We pay special attention to the intrinsic spectra of the rotated bulk structures
on two sides, \textit{i.e.} the intensities $|\hphi_{\pm}(x,\mbk)|$. 
They can be calculated from the bulk profile directly. 
It turns out that for the three pairs of spectral indices, 
$$
(-2,0)^{T}, (-1,-1)^{T}, (1,-1)^{T},
$$
and their opposites, the intensities $|\hphi_{\pm}(x,\mbk)|$ are
approximately $0.15$.
On the other hand, for the other spectra, the intensities are less than
$10^{-2}$.
This result is independent of $x$ and the tilt angle $\theta$.
For this reason, we denote the above three indices as \textbf{bulk spectral indices (BSI)}
which can be defined by
\begin{equation}
	\Lambda_{b} = \left\{ \mbk: |\hphi_{\pm}(x,\mbk)| > \varepsilon \right\},
	~~~~ \forall x\in\mbbR.
	\label{eq:BSI.def}
\end{equation}

\begin{figure}[!htbp]
	\centering
	\subfigure[$x=L$]{\label{fig:GB.spectra.16.a}
		\includegraphics[width=0.48\linewidth]{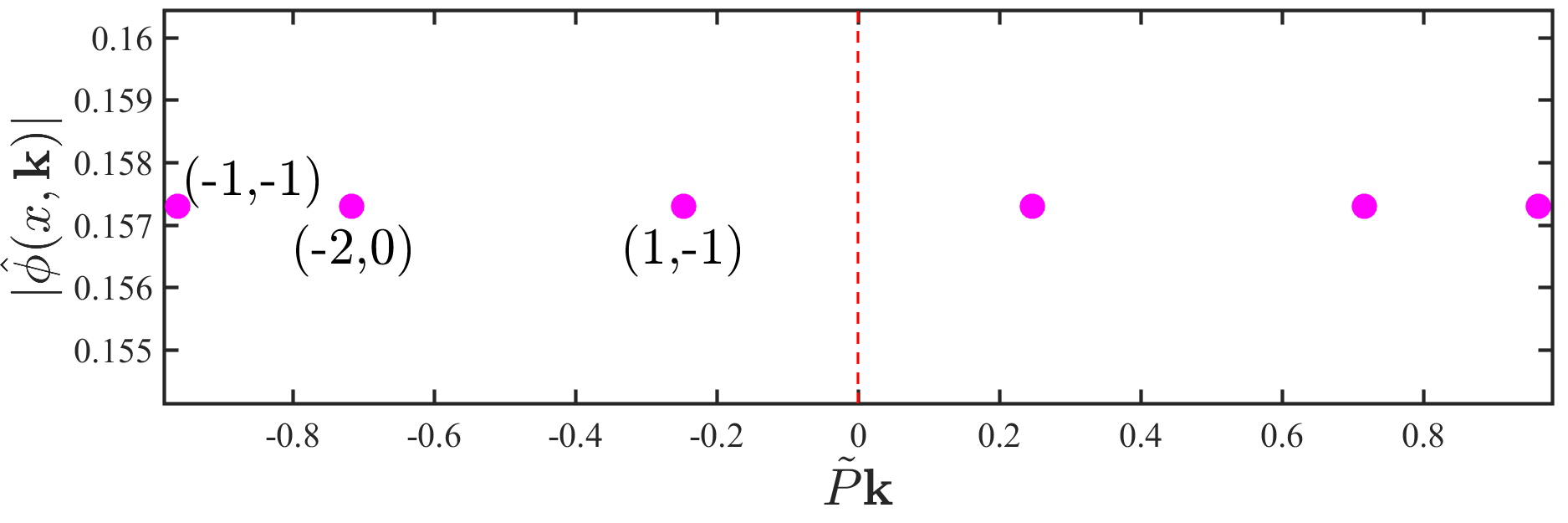}}
	\subfigure[$x=0.7686L$]{\label{fig:GB.spectra.16.b}
		\includegraphics[width=0.48\linewidth]{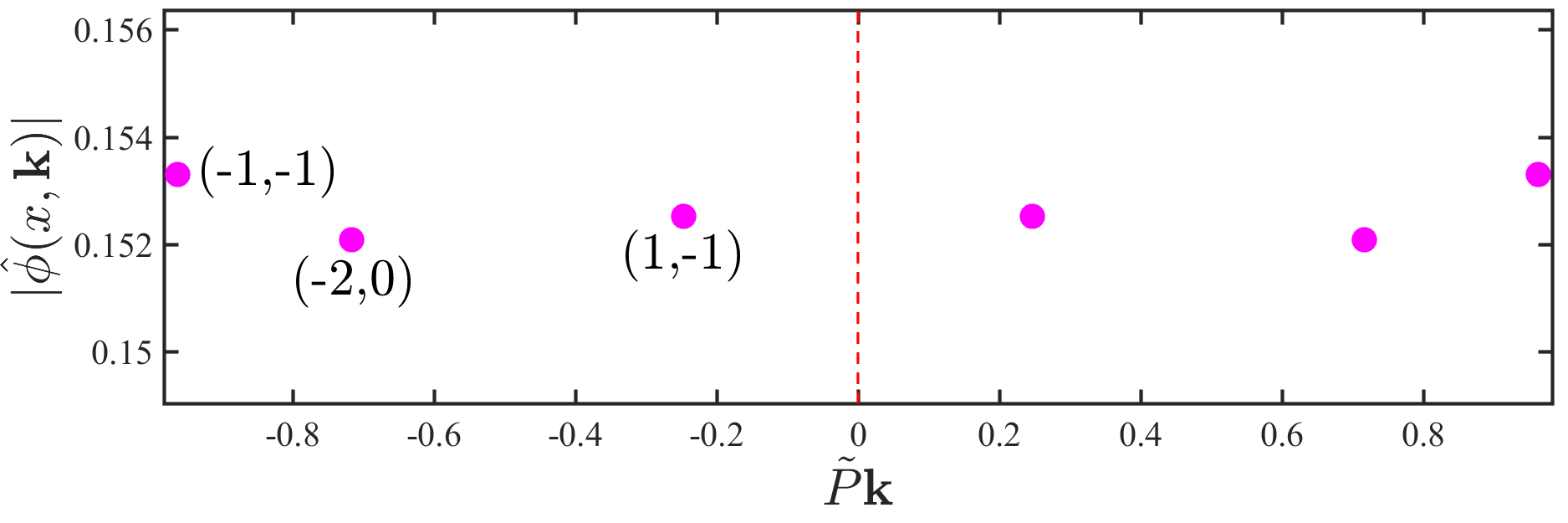}}
	\subfigure[$x=0.2605L$]{\label{fig:GB.spectra.16.c}
		\includegraphics[width=0.48\linewidth]{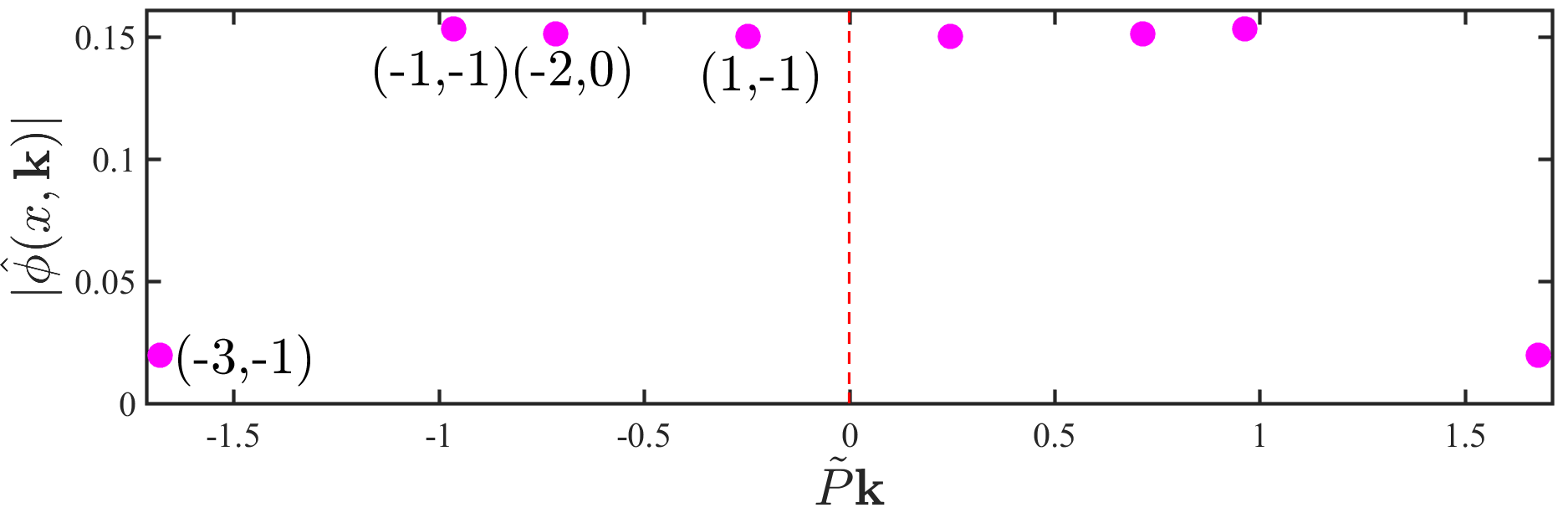}}
	\subfigure[$x=0$]{\label{fig:GB.spectra.16.d}
		\includegraphics[width=0.48\linewidth]{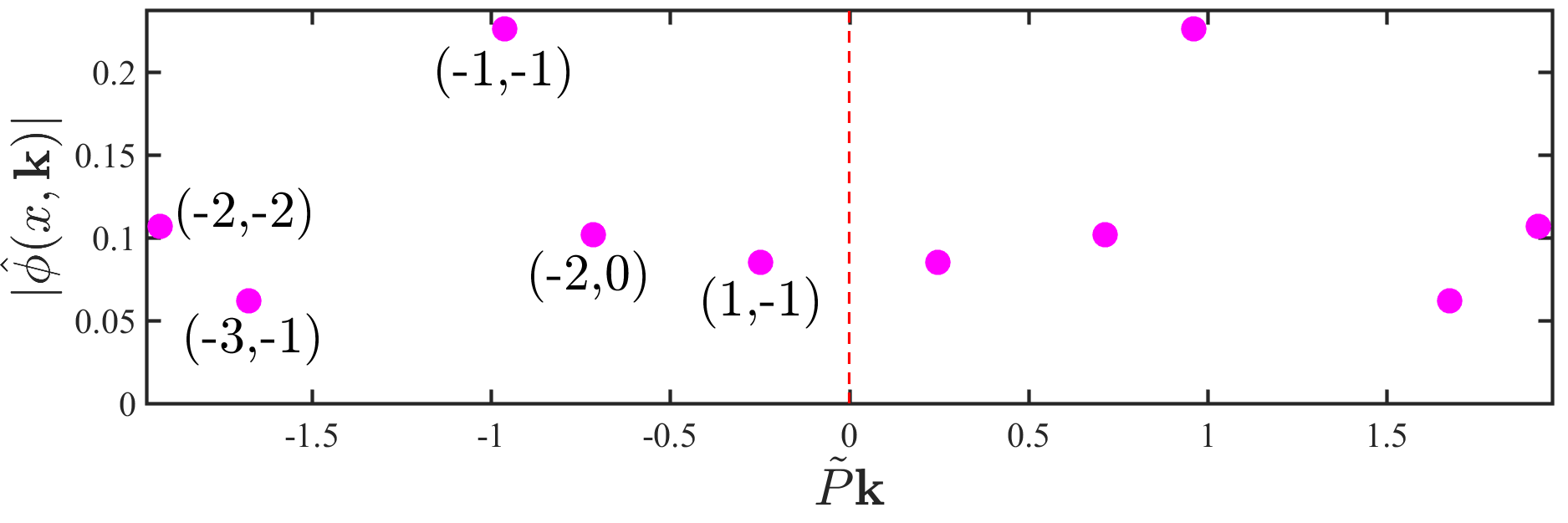}}
	\caption{\label{fig:GB.spectra.16}
		The spectral distribution on different $x$-slices whose indices belong to the PSI
		($\varepsilon = 10^{-2}$) for the hexagonal GB with the tilt angle $\theta = 16\pi/63$.
		We mark the indices $(k_{1},k_{2})^{T}$ of the spectral points on the negative side. 
		}
\end{figure}

\begin{figure}[!htbp]
	\centering
	\subfigure[Morphology with all spectra]{
		\label{fig:GB.recover.all}
		\includegraphics[width=0.47\linewidth]{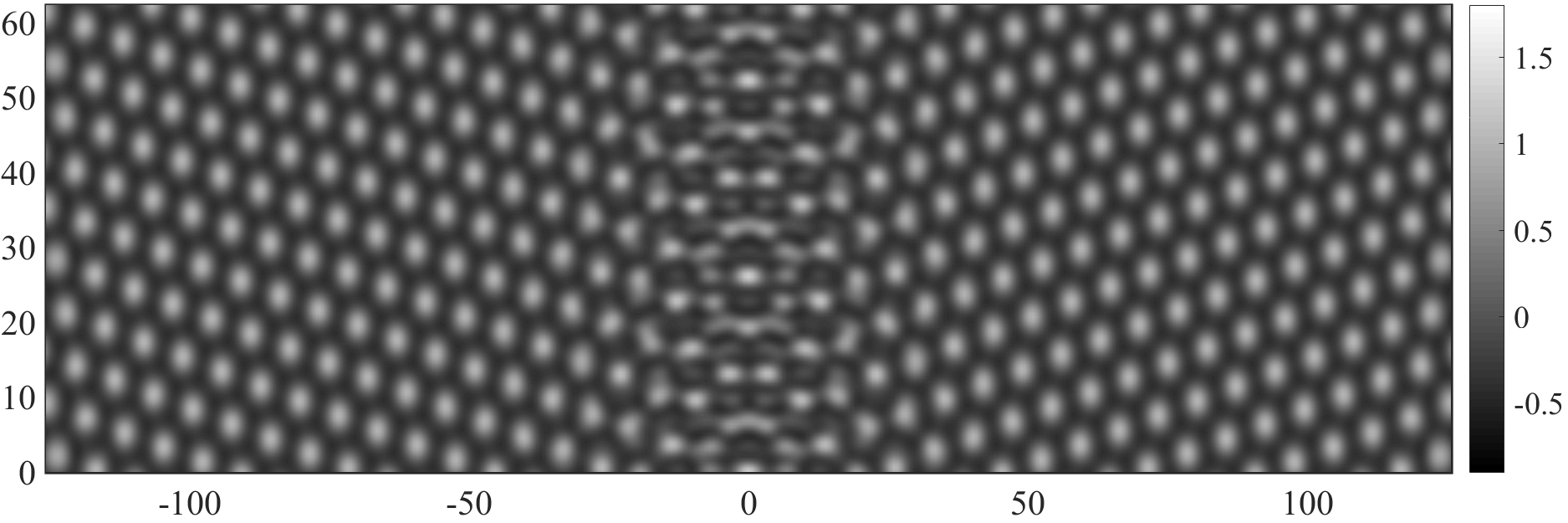}}
	\subfigure[Morphology with PSI of $\varepsilon = 10^{-2}$]{
		\label{fig:GB.recover.eps}
		\includegraphics[width=0.47\linewidth]{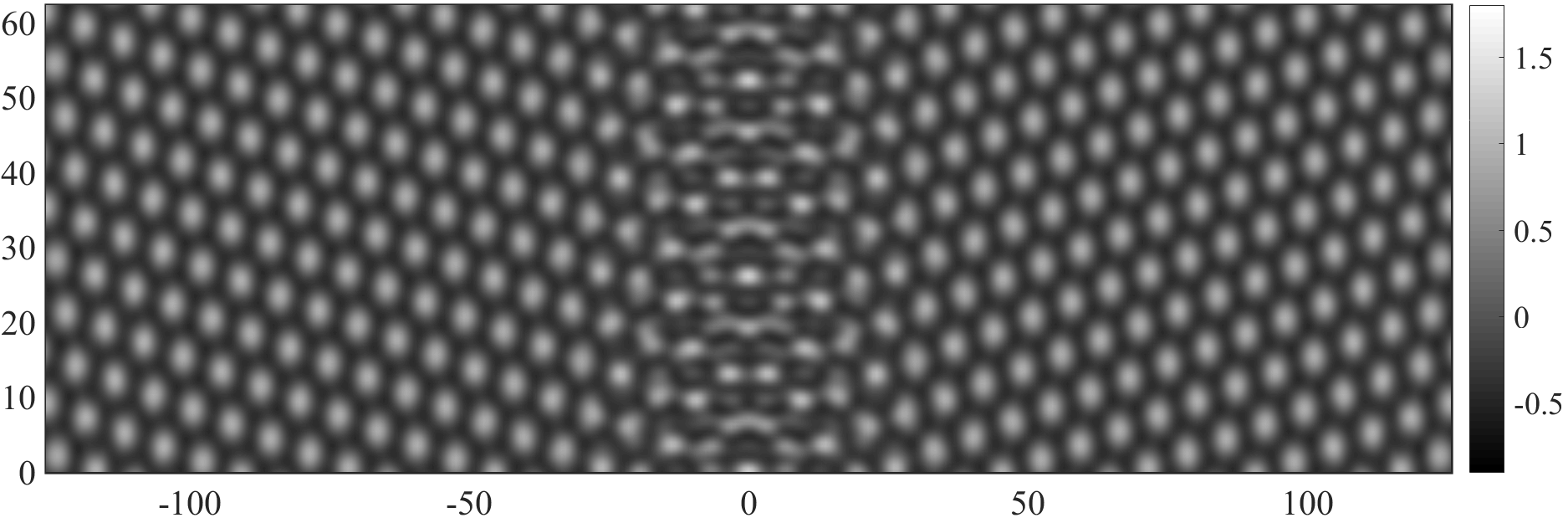}}
	\caption{\label{fig:GB.recover}
		The real-space morphology of the hexagonal GB with tilt angle
		$\theta = 16\pi/63$.
		(a) The morphology is obtained by using all spectral information.
		(b) The morphology is generated by the spectral information of PSI
		\cref{eq:PSI.def} where $\varepsilon = 10^{-2}$.
		}
\end{figure}

From the intensities of $\phi_{\pm}$, we set the threshold as
$\varepsilon=10^{-2}$ for the GBs to be examined below.
To verify the suitability of this choice, let us examine the intensities at
different $x$-slices for GBs, for which we present the tilt angle
$\theta=16\pi/63$ as an example. 
As shown in \cref{fig:GB.spectra.16}, we take some typical $x$-slices to plot
the PSI along with their intensities. 
We mark the indices $(k_{1},k_{2})^{T}$ of the spectral points on the left part.
The corresponding indices of the spectral points on the right part are 
$(-k_{1},-k_{2})^{T}$ because the spectral indices always appear in opposite pairs. 
The first one is a slice at the right anchoring plane (see
\cref{fig:GB.spectra.16.a}) where the set of spectral points is $\Lambda_{b}$.
The intensities of the three spectral points are about $0.15$. 
A slice at $x=0.7686L$ (see \cref{fig:GB.spectra.16.b}) shows that the
intensities of the BSI fluctuate very little. 
\cref{fig:GB.spectra.16.c} gives a slice closer to the center of the GB.
There appears a new spectral mode with the index $(-3,-1)^{T}$.
\cref{fig:GB.spectra.16.d} shows the spectral distribution on the plane $x=0$. 
Except for the $\Lambda_{b}$, there are two modes of indices $(-3,-1)^{T}$ and 
$(-2,-2)^{T}$.
These results confirm that the $\Lambda_{b}$ is invariant for different 
$x$-slices.
Moreover, only a couple of extra spectral modes appear to contribute to the GB profile. 
The choice $\varepsilon = 10^{-2}$ is further confirmed by comparing the
morphology of GB and that given by PSI.
They are shown in \cref{fig:GB.recover} for $\theta=16\pi/63$, where
\cref{fig:GB.recover.all} is the morphology of the GB recovered by all spectra, 
and \cref{fig:GB.recover.eps} is generated by PSI \cref{eq:PSI.def} with
$\varepsilon=10^{-2}$.
It is clear that the morphologies are nearly identical. 
In this way, we carefully examine the GBs of hexagonal crystals with different
tilt angles, and find that $\varepsilon = 10^{-2}$ is a proper value of threshold. 
A systematical analysis on the GBs of different tilt angles is presented in
\cref{tab:num.inten} showing that the intensities of most spectral modes are
less than $10^{-2}$. 

Next, we analyze the distribution of PSI on different slices. 
From the statistical results of spectral modes, we find that 
the percentages of $\Lambda_p(0)$
are all higher than that of $\Lambda_p(x)$ in total Legendre Gauss points.
Further analysis of PSI demonstrates that the PSI $\Lambda_p(0)$ on the plane
$x=0$ indeed include all spectral modes of $\Lambda_p(x)$ when $x\ne 0$, 
which can also be seen from the PSI for $\theta=16\pi/63$ in \cref{fig:GB.spectra.16}.
As a result, we shall focus on the PSI at $x=0$, and we denote
\begin{equation}
	\Lambda_{p} = \left\{ \mbk: |\hphi(0,\mbk)| > \varepsilon \right\}.
	\label{eq:PSI.redef}
\end{equation}
\begin{table}[!htbp]
	\centering
	\caption{\label{tab:num.inten}
		Statistical results of the spectral modes in the tilt GBs between
		hexagonal crystals.
		The second to fourth columns show the statistics of all spectral points,
		and the fifth to seventh columns are the results of the plane $x=0$.
		PSI \% shows the percentage of primary spectral modes
		when $\varepsilon = 10^{-2}$.
		}
	\begin{threeparttable}
		\footnotesize{
	\begin{tabular}{|c|c|c|c|c|c|c|}
		\hline
		& \multicolumn{3}{c|}{All spectra} 
		& \multicolumn{3}{c|}{Spectra on the slice of $x=0$} \\
		\hline
		$\theta$ & $|\hphi|>10^{-2}$ & $|\hphi|\leq 10^{-2}$ & PSI \%
		& $|\hphi|>10^{-2}$ & $|\hphi|\leq 10^{-2}$ & PSI \% \\
		\hline
		$11\pi/63$ & 5504 & 199696 & 2.68\% & 24 & 376 & 6.00\% \\
		\hline
		$12\pi/63$ & 4518 & 200682 & 2.20\% & 20 & 380 & 5.00\% \\
		\hline
		$13\pi/63$ & 4364 & 200836 & 2.13\% & 16 & 384 & 4.00\% \\
		\hline
		$14\pi/63$ & 4422 & 200778 & 2.16\% & 18 & 382 & 4.50\% \\
		\hline
		$15\pi/63$ & 4342 & 200858 & 2.12\% & 10 & 390 & 2.50\% \\
		\hline
		$16\pi/63$ & 4462 & 200738 & 2.17\% & 10 & 390 & 2.50\% \\
		\hline
		$17\pi/63$ & 4274 & 200926 & 2.08\% & 10 & 390 & 2.50\% \\
		\hline
		$18\pi/63$ & 4350 & 200850 & 2.12\% & 12 & 388 & 3.00\% \\
		\hline
		$19\pi/63$ & 4581\tnote{*} & 200619\tnote{*} & 2.23\% & 20 & 380 & 5.00\% \\
		\hline
	\end{tabular}
	}
	\begin{tablenotes}
		\footnotesize
	\item[*] The odd number comes from the spectral index $(0, 0)$.
	\end{tablenotes}
	\end{threeparttable}
\end{table}

\begin{figure}[!htbp]
	\centering
	\includegraphics[width=0.90\linewidth]{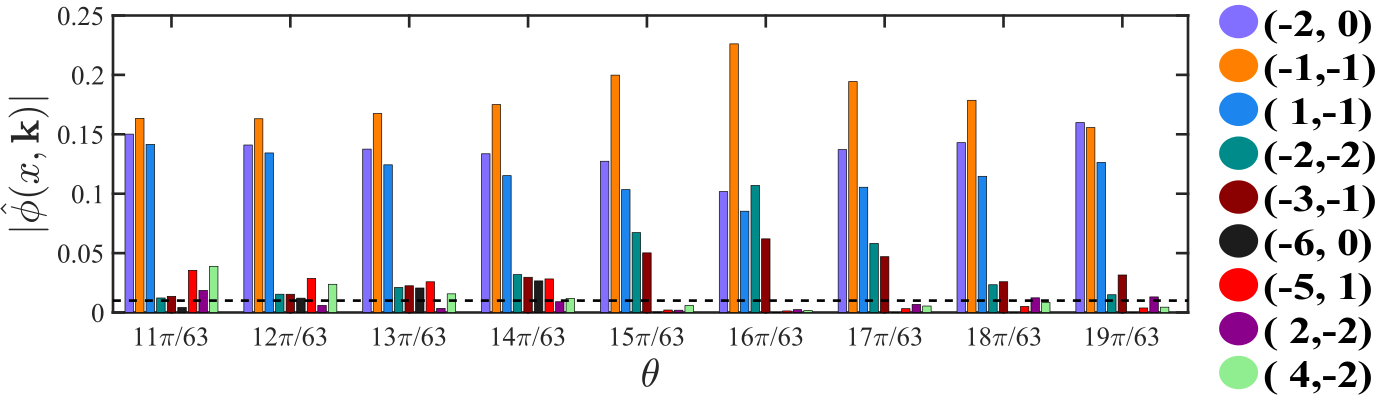}
	\caption{\label{fig:GB.PSI.intensity}
		The intensity of these spectral points with the tilt angle.
		The black dotted line corresponds to the threshold $10^{-2}$.}
\end{figure}

Among the PSI, the BSI usually have stronger intensities and make more
contributions in forming the GB structure. 
As shown in \cref{fig:GB.PSI.intensity}, we plot the intensity of some
representative spectral points under various tilt angles.
The black dotted line corresponds to the threshold $10^{-2}$.
The bar graph shows that the three bulk spectra 
(slate blue \textcolor{SlateBlue1}{$\bullet$}, 
dark orange \textcolor{DarkOrange1}{$\bullet$}, 
dodger blue \textcolor{DodgerBlue2}{$\bullet$}) 
always have intensities at the magnitude of $10^{-1}$.
In the GBs with the three tilt angles $15\pi/63$, $16\pi/63$, $17\pi/63$, the
intensities of the two spectral indices $(-2,-2)^{T}$ and $(-3,-1)^{T}$ become larger. 
For all the other spectral points, they exhibit much weaker intensities. 

It shall be pointed out that the projection method is crucial for us to extract
the few spectra of GBs, because this method provides an accurate representation
of spectra.

\subsection{Invariant representation of the PSI}
\label{subsec:represent}

To understand the relation between the intrinsic spectra of the bulk profile and
the primary spectra of the GB structure, we use the spectral indices of the BSI
to represent those of the PSI. 
Among the three BSI $\Lambda_{b} = \left\{ (-2,0)^{T}, (-1,-1)^{T}, (1,-1)^{T}
\right\}$, we choose the two linearly independent ones $(-1,-1)^{T}$ and
$(1,-1)^{T}$ to represent the PSI, 
\begin{equation}
	(k_{1}, k_{2})^{T} = j_{1} (-1, -1)^{T} + j_{2} (1, -1)^{T},
	~~~~ j_{1}, j_{2} \in \mbbZ,
	~~ (k_{1}, k_{2})^{T} \in \Lambda_{p}.
\end{equation}
The representations of the PSI with various tilt angles are listed in \cref{tab:represent}.
It is noticed that the coefficients of PSI are all integers (notice that some
spectral indices cannot be expressed with integer coefficients, such as
$(1,0)^T$). 
Moreover, the coefficients $j_{1}$ and $j_{2}$ are not greater than $5$. 
Thus the spectra can be truncated by $5$ and the GB system can be described by
very few discrete points.
For all the tilt angles, two spectral indices $(-3,-1)^{T}$ and $(-2,-2)^{T}$
always exist in the PSI. 
Of other spectral indices, the PSI has similarities when $\theta$ is relatively
small (for $\theta=l\pi/63$ where $l=11,12,13,14$).
At intermediate tilt angles $\theta=15\pi/63,16\pi/63,17\pi/63$, only
$(-3,-1)^{T}$ and $(-2,-2)^{T}$ are included in PSI in addition to BSI. 
A few spectral points appear in PSI for $18\pi/63$ and $19\pi/63$, which are
mostly distinct from those in small $\theta$. 
Despite there are differences in the PSI over angles, the number of PSI is at
most 12 (pairs), indicating that only a few spectra play a key role in the
formation of tilt GBs between hexagonal grains.
In addition, the fact that the coefficients are small integers implies that it
might be sufficient to use a discretization of a much smaller dimension to
reach reasonable accuracy for the computation of hexagonal tilt GBs.
\begin{table}[!htbp]
	\centering
	\caption{\label{tab:represent}
	  The representation of PSI by the bulk indices $(-1,-1)^{T}$ and
	  $(1,-1)^{T}$. The integers $j_{1}$ and $j_{2}$ are the coefficients before
	  the two bulk indices.
	  \Checkmark marks whether a spectral index is in the PSI. 
	}
		\footnotesize{
	\begin{tabular}{|c|c|c|c|c|c|c|c|c|c|c|c|c|c|c|}
		\hline
		$k_{1}$ & -9 & -8 & -6 & -5 & -4 & -3 & -2 & -1 &  1 &  2 &  3 &  4 &  4
		& 7 \\
		\hline
		$k_{2}$ &  1 &  2 &  0 &  1 &  0 & -1 & -2 & -3 & -3 & -2 & -3 & -4 & -2
		& -3 \\
		\hline
		$j_{1}$ &  4 &  3 &  3 &  2 &  2 &  2 &  2 &  2 &  1 &  0 &  0 &  0 & -1
		& -2 \\
		\hline
		$j_{2}$ & -5 & -5 & -3 & -3 & -2 & -1 &  0 &  1 &  2 &  2 &  3 &  4 &  3
		& 5 \\
		\hline
		$11\pi/63$ & & \Checkmark & & \Checkmark & & \Checkmark & \Checkmark & &
		\Checkmark & \Checkmark & & \Checkmark & \Checkmark & \Checkmark \\
		\hline
		$12\pi/63$ & \Checkmark & \Checkmark & \Checkmark & \Checkmark & & \Checkmark
		& \Checkmark & & & & & & \Checkmark & \\
		\hline
		$13\pi/63$ & & & \Checkmark & \Checkmark & & \Checkmark & \Checkmark & & & & & &
		\Checkmark & \\
		\hline
		$14\pi/63$ & & & \Checkmark & \Checkmark & & \Checkmark & \Checkmark & & & &
		\Checkmark & & \Checkmark & \\
		\hline
		$15\pi/63$ & & & & & & \Checkmark & \Checkmark & & & & & & & \\
		\hline
		$16\pi/63$ & & & & & & \Checkmark & \Checkmark & & & & & & & \\
		\hline
		$17\pi/63$ & & & & & & \Checkmark & \Checkmark & & & & & & & \\
		\hline
		$18\pi/63$ & & & & & & \Checkmark & \Checkmark & & & \Checkmark & & & &
		\\
		\hline
		$19\pi/63$ & & & & & \Checkmark & \Checkmark & \Checkmark & \Checkmark & 
		\Checkmark & \Checkmark & \Checkmark & & & \\
		\hline
		Times & 1 & 2 & 3 & 4 & 1 & 9 & 9 & 1 & 2 & 3 & 2 & 1 & 4 & 1 \\
		\hline
	\end{tabular}
	}
\end{table}

\begin{figure}[!htbp]
	\centering
	\includegraphics[width=0.87\linewidth]{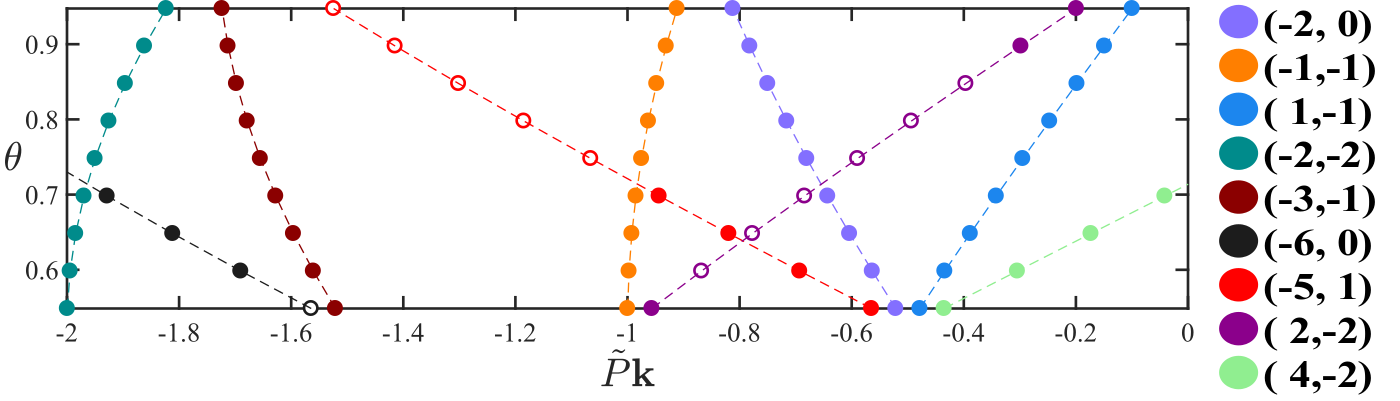}
	\caption{\label{fig:GB.PSI.spectra}
		The actual spectral points with the tilt angle.
		The spectral points whose indices belong to the PSI are marked by the
		solid dots and the others by the hollow dots.
		}
\end{figure}

More importantly, the representation of the PSI by the BSI is invariant with
various tilt angles. 
We plot in \cref{fig:GB.PSI.spectra} the spectral points of PSI for those
appearing at least three times in \cref{tab:represent}.
The spectral points marked by \Checkmark are plotted by solid dots, and the
others by hollow dots. 
The color map is given on the right side. 
The actual spectral points, determined by $\tmcP\mbk$ in
\cref{eq:actual.spectra} which are related to the tilt angle $\theta$, vary to a
great extent. 
However, their relation to the bulk spectra, given by $j_1$ and $j_2$, has commonality.

\subsection{Interface width}
\label{subsec:width}

\begin{figure}[!htbp]
  \centering
  \subfigure[Morphology by all spectra]{\label{fig:16GB.realwidth.a}
    \includegraphics[width=0.48\linewidth]{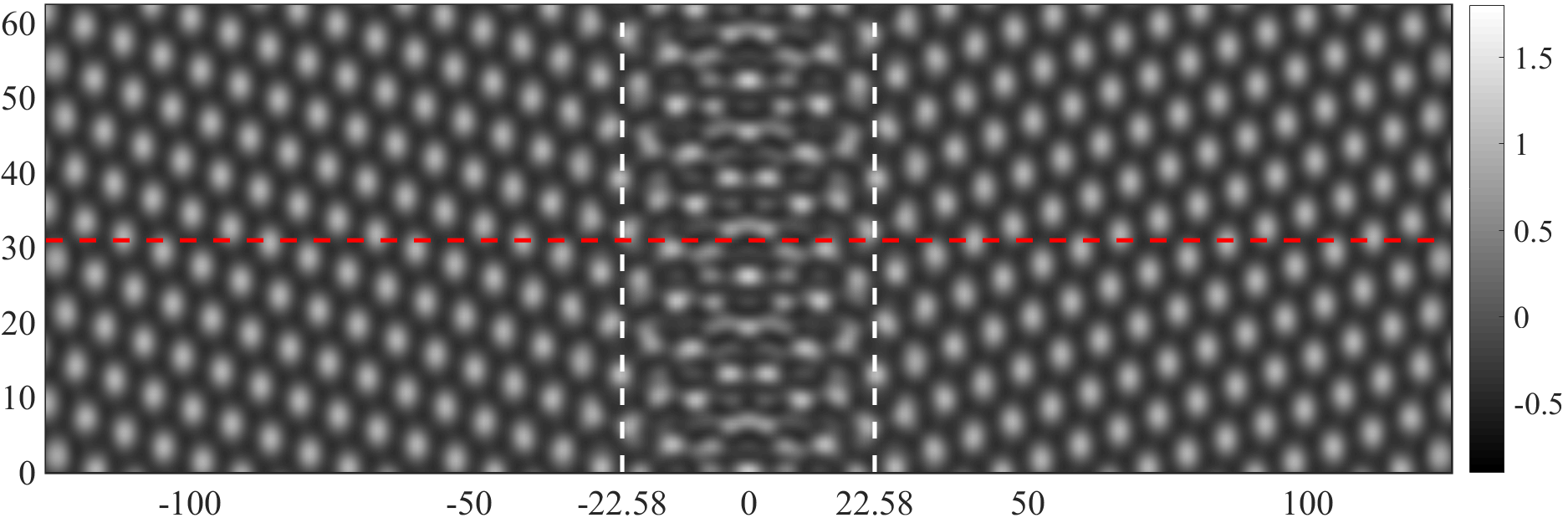}}
  \subfigure[Density distribution at $y = 10\pi$]{\label{fig:16GB.realwidth.b}
    \includegraphics[width=0.48\linewidth]{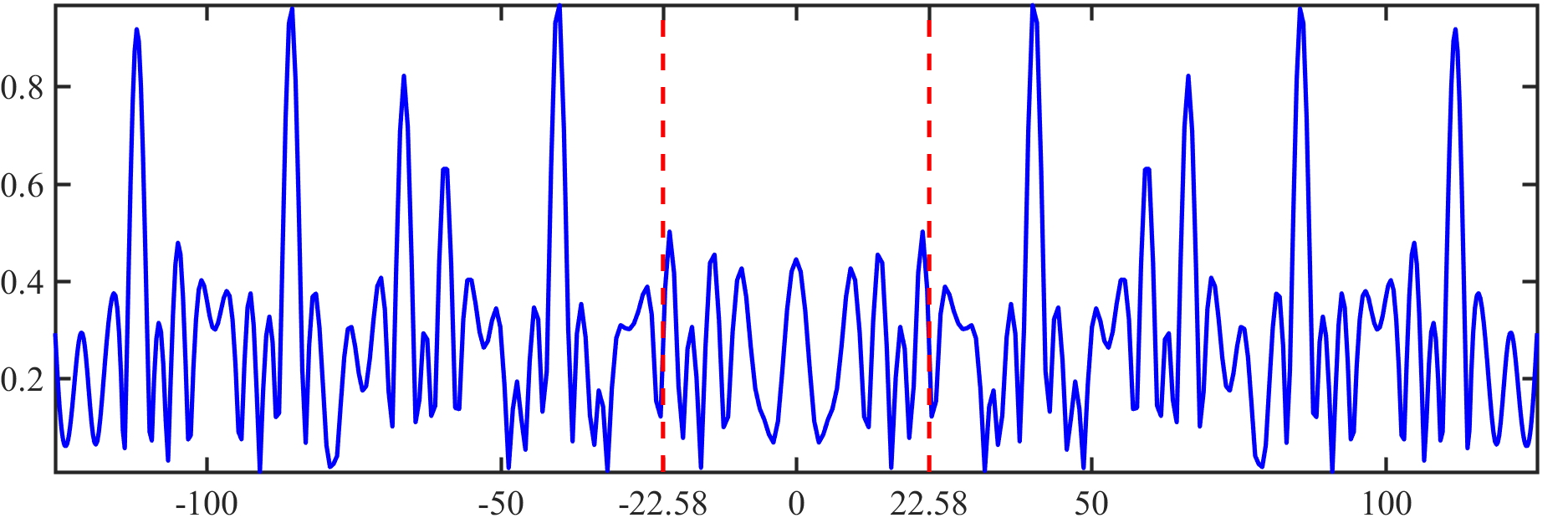}}
  \caption{\label{fig:16GB.realwidth}
    (a) The real-space morphology of the hexagonal GB with tilt angle $16\pi/63$.
    Using the spectral information, the interface edge is marked by two white
	dotted lines and its width is estimated as $45.16$.
    (b) The density distribution of the GB on the slice of $y = 10\pi$
	corresponding to the red dotted line of (a).
  }
\end{figure}

For an interface, its width is an essential parameter. 
When an interface is between two homogeneous phases, the width can be easily
extracted from the real-space profile, such as by an isosurface of the profile.  
However, this approach becomes difficult if the bulk phases are modulated.
Although in some cases the width can be roughly estimated by vision, it lacks a
unified standard, so that the comparison between different interfaces are impossible.
As an example, we look into the morphology of the tilt hexagonal GB with
$\theta=16\pi/63$, shown in \cref{fig:16GB.realwidth}.
When we investigate the density distribution for a fixed $y$, say $y=10\pi$
given in \cref{fig:16GB.realwidth.b}, we find it irregular and difficult to
propose an appropriate formula for interface width.

\begin{figure}[!htbp]
	\centering
	\subfigure[Intensity distribution of $(-2,0)^{T}$]{\label{fig:16GB.specwidth.a}
		\includegraphics[width=0.48\linewidth]{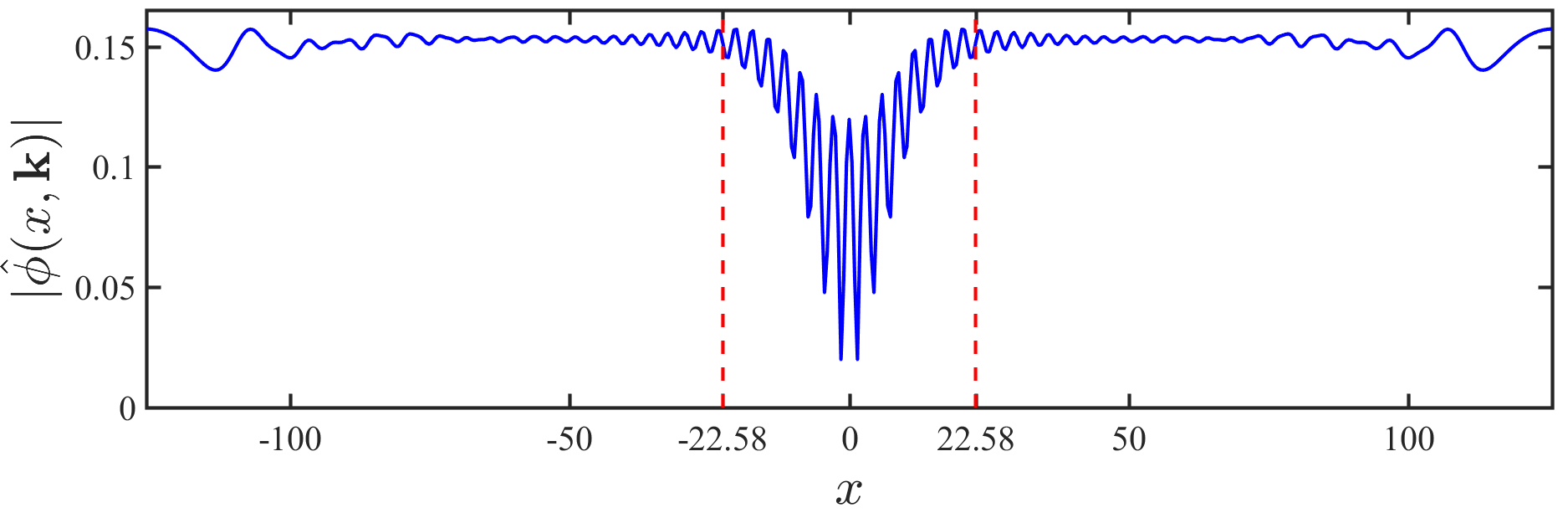}} 
	\subfigure[Intensity distribution of $(-1,-1)^{T}$]{\label{fig:16GB.specwidth.b}
		\includegraphics[width=0.48\linewidth]{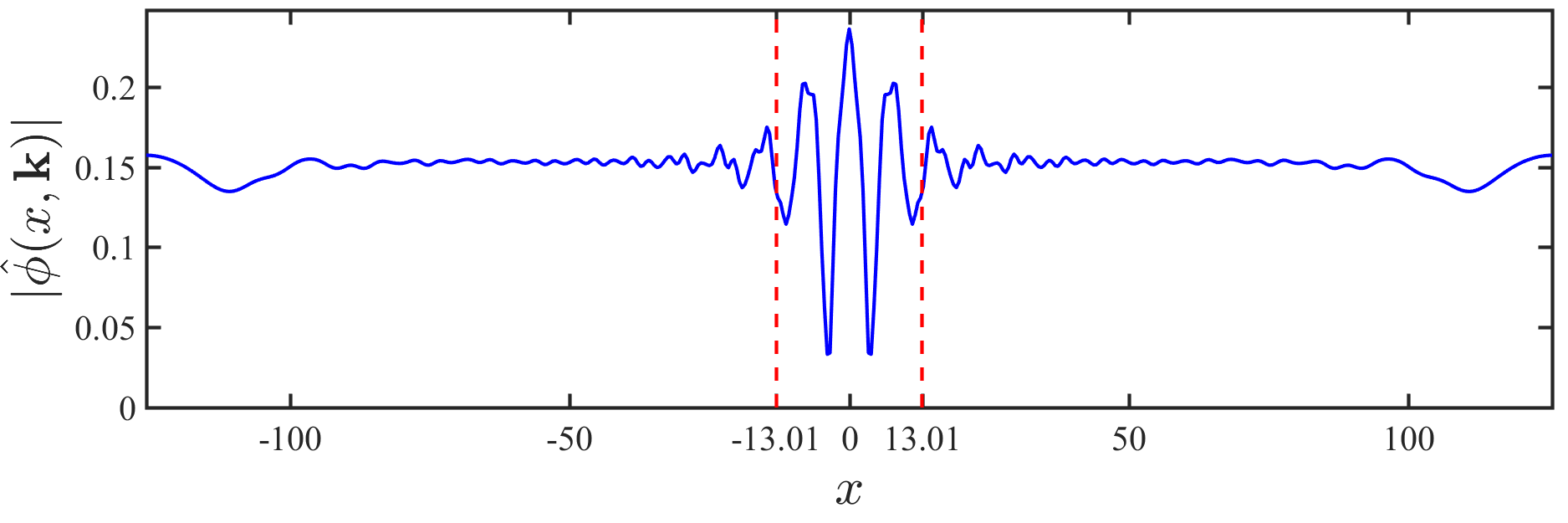}} 
	\subfigure[Intensity distribution of $(1,-1)^{T}$]{\label{fig:16GB.specwidth.c}
		\includegraphics[width=0.48\linewidth]{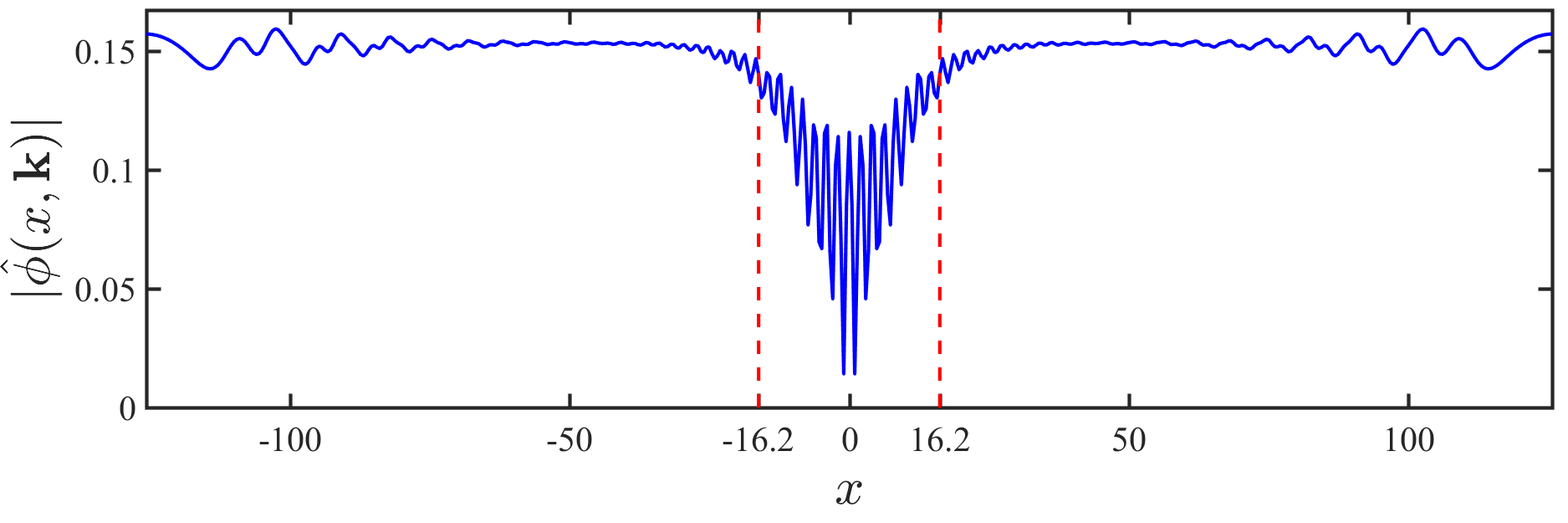}} 
	\caption{\label{fig:16GB.specwidth}
		The intensity of the bulk spectra of the BSI along the $x$-direction in
		the GB of hexagonal crystals with tilt angle $16\pi/63$.
		Their oscillation widths are (a) $I((-2,0)^{T}) = [-22.58, 22.58]$; (b)
		$I((-1,-1)^{T}) = [-13.01, 13.01]$; (c) $I((1,-1)^{T}) = [-16.20, 16.20]$.
		Thus the interface region is $I = [-22.58, 22.58]$ and its width is $45.16$.}
\end{figure}

In contrast, when we examine the GB in the spectral space, the transition
between the bulk region and the interface region become much clearer. 
We propose to use the BSI to define the interface width, for which we still take
the GB with the tilt angle $16\pi/63$ as an example to illustrate.
We plot the intensities of BSI with respect to $x$ (see \cref{fig:16GB.specwidth}\,(a-c)).
We observe that the intensities exhibit a larger oscillation close to $x=0$,
while much smaller far away from $x = 0$. 
Therefore, we define the interface width from the oscillation of the bulk spectra of the BSI.
First, let us define the oscillation of the spectral index $\mbk$, 
\begin{equation}
	H(x_{k},\mbk) = \left||\hphi(x_{1},\mbk)| - |\hphi(x_{2},\mbk)|\right|,
	~~~~ x_{k} = (x_{1}+x_{2})/2,
	\label{eq:define.hk}
\end{equation}
where $|\hphi|$ is the modulus of $\hphi$.
$x_{1}$ and $x_{2}$ are two adjacent stationary points satisfying
$\partial|\phi(x_{j},\mbk)| / \partial x_{j} = 0, ~ j = 1, 2$.
Then, we could define the left oscillation edge of the spectral index $\mbk$, 
$I_{-}(\mbk)$, and the right one $I_{+}(\mbk)$, 
\begin{equation}
	\left\{
	\begin{aligned}
		I_{-}(\mbk) &= \argmax_{x_{k}} x_{k}, ~~ x_{k} < 0, \\
		I_{+}(\mbk) &= \argmin_{x_{k}} x_{k}, ~~ x_{k} > 0, \\
	\end{aligned}
	\right.
	~~ \textit{s.t.} ~~ 
	H(x_{k},\mbk) < \rho_{H}\max_{x}\{H(x,\mbk)\},
\end{equation}
where $\rho_{H}\in(0,1)$ is a constant. 
Based on our investigation of the intensities, we choose $\rho_{H} = 0.1$. 
Denote $I(\mbk) = [I_{-}(\mbk), I_{+}(\mbk)]$. 
The interface region is defined by 
\begin{equation}
	I = I((-2,0)^{T}) \cup I((-1,-1)^{T}) \cup I((1,-1)^{T}).
	\label{eq:def.inter.reg}
\end{equation}
The interface width is computed by the length of $I$. 

With the definition of interface width, we mark the oscillation edge of the bulk
spectra of the BSI in \cref{fig:16GB.specwidth}\,(a-c).
The interface region defined by \cref{eq:def.inter.reg} is $[-22.58, 22.58]$
which is consistent with the estimation by vision. 
Furthermore, we compute the interface width in the tilt GB of hexagonal phases
with various tilt angles (see \cref{tab:GB.width.cases}).
In the GB with tilt angle $18\pi/63$ or $19\pi/63$, the interface region is not
mirror-symmetric about the $x=0$. 
This is possible as mirror-symmetric profiles may be unstable.
\begin{table}[!htbp]
	\centering
	\caption{\label{tab:GB.width.cases}
		The interface region defined by \cref{eq:def.inter.reg} in the GBs
		of hexagonal phases with various tilt angles.
		}
\footnotesize{
	\begin{tabular}{|c|c|c|c|c|c|}
		\hline
		tilt angle & $I((-2,0)^{T})$ & $I((-1,-1)^{T})$ & $I((1,-1)^{T})$ & $I$
		& width \\
		\hline
		$11\pi/63$ & $[-41.97, 41.97]$ & $[-30.19, 30.19]$ & $[-44.18, 44.18]$ & 
			$[-44.18, 44.18]$ & $88.36$ \\
		\hline
		$12\pi/63$ & $[-16.20, 16.20]$ & $[-34.85, 34.85]$ & $[-18.65, 18.65]$ & 
			$[-34.85, 34.85]$ & $69.70$ \\
		\hline
		$13\pi/63$ & $[-14.48, 14.48]$ & $[-23.07, 23.07]$ & $[-17.18, 17.18]$ &
			$[-23.07, 23.07]$ & $46.14$ \\
		\hline
		$14\pi/63$ & $[-14.97, 14.97]$ & $[-17.18, 17.18]$ & $[-16.69, 16.69]$ &
			$[-17.18, 17.18]$ & $34.36$ \\
		\hline
		$15\pi/63$ & $[-21.35, 21.35]$ & $[-0.74,  0.74]$ & $[-16.44, 16.44]$ &
			$[-21.35, 21.35]$ & $42.71$ \\
		\hline
		$16\pi/63$ & $[-22.58, 22.58]$ & $[-13.01, 13.01]$ & $[-16.20, 16.20]$ &
			$[-22.58, 22.58]$ & $45.16$ \\
		\hline
		$17\pi/63$ & $[-15.95, 15.95]$ & $[-6.63,  6.63]$ & $[-15.71, 15.71]$ &
			$[-15.95, 15.95]$ & $31.91$ \\
		\hline
		$18\pi/63$ & $[-15.46, 9.08]$ & $[-9.57,  3.44]$ & $[-19.88, 13.74]$ &
			$[-19.88, 13.74]$ & $33.62$ \\
		\hline
		$19\pi/63$ & $[-6.63, 16.94]$ & $[-0.74, 10.55]$ & $[-10.30, 24.05]$ &
			$[-10.30, 24.05]$ & $34.36$ \\
		\hline
	\end{tabular}
}
\end{table}
The corresponding real-space morphologies can be found in
\cref{fig:GB.realwidth.cases}, where the interface edges calculated from
spectral information are marked by two white dotted lines. 
It should be noted that the tilt GB of hexagonal phases with a small tilt angle
($11\pi/63$, $12\pi/63$, $13\pi/63$) is difficult to distinguish by vision. 
But the interface region \cref{eq:def.inter.reg} defined from the spectral
perspective provides an appropriate standard. 
For the others in \cref{fig:GB.realwidth.cases}, the interface edge is almost
consistent with the estimate given by vision.

\section{Conclusion}
\label{sec:conclu}

We provide a new understanding of tilt GBs between hexagonal grains from the
perspective of spectra. 
The GBs are regarded as structures filling an infinite banded region between two
parallel planes.
Using the projection method to deal with quasiperiodicity, we can accurately
capture the spectra along the direction of the dividing planes. 

We find that a few spectral modes effectively contribute to the formation of
tilt GBs between hexagonal grains. 
The capture of these constituents is a fine consequence of using the projection
methods to accurately represent the spectral points. 
These spectral points turn out to have invariant indices for different tilt angles. 
From the spectral information, we are able to propose a good definition of the
interface width from the oscillation of the intensities.
The resulting widths are consistent with the estimate by vision. 

Our results demonstrate that the spectral viewpoint is able to reveal several
ingredients of GBs that are not easily accessible from the real-space profile. 
Furthermore, the results imply that very few discrete spectral points might be
sufficient to describe the essential features of the tilt GBs, which is helpful
for GBs between other bulk phases.
These advantages are all built on the delicate formulation of the interface
system using the projection method. 
In future work, we aim to use the framework to investigate 
interfaces involving other phases from a spectral viewpoint, especially the
three-dimensional phases, including the bcc/fcc spherical and gyroid that are
periodic, and icosahedral quasicrystals.

%
%
%
%
%
%
%
%
%


\begin{thebibliography}{10}

\bibitem{aguirre2019molecular}
{\sc R.~Aguirre, S.~Abdullah, X.~Zhou, and D.~Zubia}, {\em Molecular dynamics
  calculations of grain boundary mobility in cdte}, Nanomaterials, 9 (2019),
  p.~552.

\bibitem{beyerlein2010statistical}
{\sc I.~Beyerlein, L.~Capolungo, P.~Marshall, R.~McCabe, and C.~Tom\'e}, {\em
  Statistical analyses of deformation twinning in magnesium}, Philos. Mag., 90
  (2010), pp.~2161--2190.

\bibitem{buban2006grain}
{\sc J.~P. Buban, K.~Matsunaga, J.~Chen, N.~Shibata, W.~Y. Ching, T.~Yamamoto,
  and Y.~Ikuhara}, {\em Grain boundary strengthening in {Alumina} by rare earth
  impurities}, Science, 311 (2006), pp.~212--215.

\bibitem{cao2021computing}
{\sc D.~Cao, J.~Shen, and J.~Xu}, {\em Computing interface with
  quasiperiodicity}, J. Comput. Phys., 424 (2021), p.~109863.

\bibitem{corduneanu1989almost}
{\sc C.~Corduneanu}, {\em Almost Periodic Functions}, 2$^{nd}$ edition, Chelsea
  Publishing Company, New York.

\bibitem{edwards1993parametrically}
{\sc W.~S. Edwards and S.~Fauve}, {\em Parametrically excited quasicrystalline
  surface waves}, Phys. Rev. E, 47 (1993).

\bibitem{feng2021visualizing}
{\sc X.~Feng, M.~Zhuo, H.~Guo, and E.~L. Thomas}, {\em Visualizing the
  double-gyroid twin}, PNAS, 118 (2021), pp.~1--6.

\bibitem{fischer2011colloidal}
{\sc S.~Fischer, A.~Exner, K.~Zielske, J.~Perlich, S.~Deloudi, W.~Steurer,
  P.~Lindner, and S.~Forster}, {\em Colloidal quasicrystals with 12-fold and
  18-fold diffraction symmetry}, PNAS, 108 (2011), pp.~1810--1814.

\bibitem{flint2019phasefield}
{\sc T.~F. Flint, Y.~L. Sun, Q.~Xiong, M.~C. Smith, and J.~A. Francis}, {\em
  Phase-field simulation of grain boundary evolution in microstructures
  containing second-phase particles with heterogeneous thermal properties},
  Sci. Rep., 9 (2019), p.~18426.

\bibitem{brainBOOKgrain}
{\sc G.~Gottstein and L.~S. Shvindlerman}, {\em Grain Boundary Migration in
  Metals: Thermodynamics, Kinetics, Applications}, 2$^{nd}$ edition, Taylor and
  Francis Group.

\bibitem{holmes1990poincare}
{\sc P.~Holmes}, {\em Poincar\'e, celestial mechanics, dynamical systems theory
  and "chaos"}, Phys. Rep., 193 (1990), pp.~137--163.

\bibitem{jiang2021on}
{\sc K.~Jiang, S.~Li, and P.~Zhang}, {\em On the approximation of quasiperiodic
  functions by periodic functions}, In preparation,  (2021).

\bibitem{jiang2020efficient}
{\sc K.~Jiang, W.~Si, C.~Chen, and C.~Bao}, {\em Efficient numerical methods
  for computing the stationary states of phase field crystal models}, SIAM J.
  Sci. Comput., 42 (2020), pp.~B1350--B1377.

\bibitem{jiang2015stability}
{\sc K.~Jiang, J.~Tong, P.~Zhang, and A.-C. Shi}, {\em Stability of
  two-dimensional soft quasicrystals}, Phys. Rev. E, 92 (2015), p.~042159.

\bibitem{jiang2014numerical}
{\sc K.~Jiang and P.~Zhang}, {\em Numerical methods for quasicrystals}, J.
  Comput. Phys., 256 (2014), pp.~428--440.

\bibitem{jiang2018numerical}
{\sc K.~Jiang and P.~Zhang}, {\em Numerical mathematics of quasicrystals},
  Proc. Int. Cong. of Math., 3 (2018), pp.~3575--3594.

\bibitem{li2019sculpted}
{\sc X.~Li, J.~A. Mart\'{i}nez-Gonz\'{a}lez, O.~Guzm\'{a}n, X.~Ma, K.~Park,
  C.~Zhou, Y.~Kambe, H.~M. Jin, J.~A. Dolan, P.~F. Nealey, and J.~J. de~Pablo},
  {\em Sculpted grain boundaries in soft crystals}, Sci. Adv., 5 (2019),
  p.~eaax9112.

\bibitem{lifshitz1997theoretical}
{\sc R.~Lifshitz and D.~M. Petrich}, {\em Theoretical model for faraday waves
  with multiple-frequency forcing}, Phys. Rev. Lett., 79 (1997),
  pp.~1261--1264.

\bibitem{madadi2021coarse}
{\sc A.~A. Madadi and A.~Khoei}, {\em A coarse-grained -- atomistic multi-scale
  method to study the mechanical behavior of heterogeneous {FCC}
  nano-materials}, Comput. Mater. Sci., 199 (2021), p.~110725.

\bibitem{mason2015grain}
{\sc J.~K. Mason}, {\em Grain boundary energy and curvature in monte carlo and
  cellular automata simulations of grain boundary motion}, Acta Mater., 94
  (2015), pp.~162--171.

\bibitem{ogata2005energy}
{\sc S.~Ogata, J.~Li, and S.~Yip}, {\em Energy landscape of deformation
  twinning in {BCC} and {FCC} metals}, Phys. Rev. B, 71 (2005), p.~224102.

\bibitem{riet2021molecular}
{\sc A.~A. Riet, J.~A.~V. Orman, and D.~J. Lacks}, {\em A molecular dynamics
  study of grain boundary diffusion in {MgO}}, Geochim. Cosmochim. Acta, 292
  (2021), pp.~203--216.

\bibitem{Shechtman1984}
{\sc D.~Shechtman, I.~Blech, D.~Gratias, and J.~W. Cahn}, {\em Metallic phase
  with long-range orientational order and no translational symmetry}, Phys.
  Rev. Lett., 53 (1984), pp.~1951--1953.

\bibitem{shimada2002optimization}
{\sc M.~Shimada, H.~Kokawa, Z.~J. Wang, Y.~S. Sato, and I.~Karibe}, {\em
  Optimization of grain boundary character distribution for intergranular
  corrosion resistant 304 stainless steel by twin-induced grain boundary
  engineering}, Acta Mater., 50 (2002), pp.~2331--41.

\bibitem{son2020twodimensional}
{\sc Y.~Son, H.~B. Chung, and S.~Lee}, {\em A two-dimensional monte carlo model
  for pore densification in a bi-crystal via grain boundary diffusion: Effect
  of diffusion rate, initial pore distance, temperature, boundary energy and
  number of pores}, J. Eur. Ceram. Soc., 40 (2020), pp.~3158--3171.

\bibitem{uberuaga2013point}
{\sc B.~P. Uberuaga, X.-M. Bai, P.~P. Dholabhai, N.~Moore, and D.~M. Duffy},
  {\em Point defect-grain boundary interactions in mgo: an atomistic study}, J.
  Phys.: Condens. Matter, 25 (2013), p.~355001.

\bibitem{watanabe2004toughening}
{\sc T.~Watanabe and S.~Tsurekawa}, {\em Toughening of brittle materials by
  grain boundary engineering}, Mater. Sci. Eng., A: Struct. Mater.: Properties,
  Microstruct. Process., A387--A389 (2004), pp.~447--55.

\bibitem{xu2017computing}
{\sc J.~Xu, C.~Wang, A.-C. Shi, and P.~Zhang}, {\em Computing optimal
  interfacial structure of modulated phases}, Commun. Comput. Phys., 21 (2017),
  pp.~1--15.

\bibitem{yamanaka2017phase}
{\sc A.~Yamanaka, K.~McReynolds, and P.~W. Voorhees}, {\em Phase field crystal
  simulation of grain boundary motion, grain rotation and dislocation reactions
  in a {BCC} bicrystal}, Acta Mater., 133 (2017), pp.~160--171.

\end{thebibliography}

\end{document}